\documentclass[prx,twocolumn,notitlepage,showpacs,prl,amsmath,amstex,amssymb,citeautoscript,longbibliography]{revtex4-1}
\pdfoutput=1
\usepackage{natbib}
\usepackage[english]{babel}
\usepackage{letltxmacro}
\usepackage{latexsym}
\LetLtxMacro{\ORIGselectlanguage}{\selectlanguage}
\makeatletter
\DeclareRobustCommand{\selectlanguage}[1]{%
  \@ifundefined{alias@\string#1}
    {\ORIGselectlanguage{#1}}
    {\begingroup\edef\x{\endgroup
       \noexpand\ORIGselectlanguage{\@nameuse{alias@#1}}}\x}%
}
\newcommand{\definelanguagealias}[2]{%
  \@namedef{alias@#1}{#2}%
}
\makeatother
\definelanguagealias{en}{english}
\definelanguagealias{English}{english}
\usepackage{graphicx}
\usepackage{amsmath}
\usepackage{amsfonts}
\usepackage{amssymb,bbding}
\usepackage{bm}
\usepackage{color}
\usepackage[percent]{overpic}
\usepackage{soul} 
\usepackage{amssymb}
\usepackage{wasysym}
\usepackage{dsfont}
\usepackage{float}
\usepackage{braket}

\usepackage{hyperref}
\hypersetup{
    bookmarks=false,         % show bookmarks bar?
    unicode=false,          % non-Latin characters in AcrobatÕs bookmarks
    pdftoolbar=false,        % show AcrobatÕs toolbar?
    pdfmenubar=true,        % show AcrobatÕs menu?
    pdffitwindow=false,     % window fit to page when opened
    pdfstartview={FitH},    % fits the width of the page to the window
    pdftitle={},    % title
    pdfauthor={},     % author
    pdfsubject={},   % subject of the document
    pdfcreator={},   % creator of the document
    pdfproducer={}, % producer of the document
    pdfkeywords={quantum scars} {non-ergodic dynamics}, % list of keywords
    pdfnewwindow=true,      % links in new window
    colorlinks=true,       % false: boxed links; true: colored links
    linkcolor=black,          % color of internal links (change box color with linkbordercolor)
    citecolor=blue,        % color of links to bibliography
    filecolor=magenta,      % color of file links
    urlcolor=blue           % color of external links
}

\newcommand{\footnoteremember}[2]{
\footnote{#2}
\newcounter{#1}
\setcounter{#1}{\value{footnote}}
}
\newcommand{\footnoterecall}[1]{\footnotemark[\value{#1}]}

\newcommand{\be}{\begin{equation}}
\newcommand{\ee}{\end{equation}}
\newcommand{\bea}{\begin{eqnarray}}
\newcommand{\eea}{\end{eqnarray}}

\newcommand{\corr}[1]{\langle{ #1}\rangle}

\begin{document}
\title{Quantum many-body scars}
\author{C. J. Turner$^1$, A. A. Michailidis$^1$, D. A. Abanin$^2$, M. Serbyn$^3$, and Z. Papi\'c$^1$}
\affiliation{$^1$School of Physics and Astronomy, University of Leeds, Leeds LS2 9JT, United Kingdom}
\affiliation{$^2$Department of Theoretical Physics, University of Geneva,
24 quai Ernest-Ansermet, 1211 Geneva, Switzerland}
\affiliation{$^3$IST Austria, Am Campus 1, 3400 Klosterneuburg, Austria}

\date{\today}
\begin{abstract}
Certain wave functions of non-interacting quantum chaotic systems can exhibit  ``scars" in the fabric of their real-space density profile. Quantum scarred wave functions concentrate in the vicinity of unstable periodic classical trajectories. We introduce the notion of \emph{many-body} quantum scars which reflect the existence of a subset of special many-body eigenstates concentrated in certain parts of the Hilbert space. We demonstrate the existence of scars in the Fibonacci chain -- the one-dimensional model with a constrained local Hilbert space realized in the 51 Rydberg atom quantum simulator [H. Bernien \emph{et al}., arXiv:1707.04344]. The quantum scarred eigenstates are embedded throughout the thermalizing many-body spectrum, but surprisingly lead to direct experimental signatures such as robust oscillations following a quench from a charge-density wave state found in experiment. We develop a model based on a single particle hopping on the Hilbert space graph, which quantitatively captures the scarred wave functions up to large systems of $L=32$ atoms. Our results suggest that scarred many-body bands give rise to a new universality class of quantum dynamics, which opens up opportunities for creating and manipulating novel states with long-lived coherence in systems that are now amenable to experimental study.
\end{abstract}
\maketitle

\emph{Introduction.---} Controllable, quantum-coherent systems of ultracold atoms~\cite{Kinoshita06, Bloch15}, trapped ions~\cite{Monroe16}, and nitrogen-vacancy spins in diamond~\cite{Lukin16} have emerged as platforms for realising and probing highly non-equilibrium quantum matter. In particular, these systems have opened the door to the investigation of \emph{non-ergodic} dynamics in isolated quantum systems. Such unusual kind of dynamics is now known to occur when there is an emergence of extensively many conserved quantities. This can happen either because of integrability conditions in low-dimensional systems when interactions are finely tuned~\cite{Sutherland}, or more generically because of the presence of quenched disorder giving rise to many-body localization (MBL)~\cite{Basko06,Serbyn13-1,Huse13}. In both cases, the system strongly violates the ``eigenstate thermalization hypothesis" (ETH)~\cite{DeutschETH,SrednickiETH}, which was conjectured to govern the properties of ergodic systems and their approach to thermal equilibrium. 

MBL and integrable systems break ergodicity in a strong way: in both cases there is an extensive number of operators that commute with the Hamiltonian, which dramatically changes their energy level statistics from Wigner-Dyson to Poisson distribution~\cite{BerryTabor,PalHuse}. This motivates the question: are there systems which only \emph{weakly} break ergodicity? In particular, are there systems in which \emph{some} eigenstates are atypical and dynamics strongly depends on the initial conditions? The existing theoretical studies, which tested the ETH numerically in systems of spins, fermions, and bosons in 1D and 2D systems~\cite{Rigol07,RigolNature} seem to rule out such a possibility. In particular, Ref.~\cite{Huse14} found that in an ergodic spin chain, {\it all} highly excited states were typical and thermal, obeying the strong version of the ETH.  

In this paper  we demonstrate that weak breaking of ergodicity can occur in kinetically-constrained 1D models which can be viewed as effective models of anyon excitations in 2D topological phases of matter, like fractional quantum Hall states and spin liquids. Topological order in these systems is connected with emergent gauge fields, which strongly modify the Hilbert space structure of the effective model that can no longer be expressed as a simple tensor product of qubits. While such models have been theoretically investigated~\cite{Feiguin07,Trebst2008,  Lesanovsky2012, Lindner2012, Glaetzle2014, Vasseur2015, Chandran16, Lan2017, Lan2017_2}, recent works~\cite{Schauss2012,Labuhn2016,Bernien17} demonstrate that they can also be realized in experiment  with  Rydberg atoms in 1D or 2D traps. We focus on the simple example of a one-dimensional chain (with the Hamiltonian defined in Eq.~(\ref{Eq:Ham}) below) whose constrained Hilbert space grows as the Fibonacci sequence, and therefore below we refer to this system as a ``Fibonacci chain". The 2D physical system corresponding to such a chain is the $\nu=\frac{12}{5}$ fractional quantum Hall state, whose quasiparticle excitations also grow in number according to the Fibonacci sequence~\cite{ReadRezayi}, and therefore support ``universal topological quantum computation"~\cite{NayakRMP}. 

Our results for the non-ergodic dynamics in the Fibonacci chain  can be summarized as follows. First, based on the energy level statistics , we find that the model exhibits level repulsion and is non-integrable. This is also supported by the ballistic spreading of entanglement entropy. Second, we show that the model has a band of \emph{special} eigenstates  coexisting with thermalizing eigenstates in the middle of the many-body band. 
The number of special eigenstates scales linearly with the system size, in contrast to the exponentially growing size of the Hilbert space. Surprisingly, even though the special eigenstates only comprise a vanishing fraction of all states in the thermodynamic limit, they have direct physical manifestations, and they can be accessed by preparing  the system in specific product states. In particular, the band of special eigenstates underlies the unexpected long-time oscillations observed  experimentally~\cite{Bernien17}. Finally, we shed light on the structure of special eigenstates by introducing an effective tight-binding method. This method allows us to obtain accurate numerical approximations of special eigenstates by solving the problem of a single particle hopping on a Hilbert space graph. Despite the apparent simplicity of this representation, special eigenstates have complicated entanglement structures which, unlike the states considered in Ref.~\cite{Bernevig2017}, cannot be expressed as matrix product states with finite bond dimension. 

The existence of a band of special eigenstates is strongly reminiscent of the phenomenon of quantum scars in single-particle chaotic billiards~\cite{Heller84}, which have been observed in microwave cavities~\cite{Sridhar1991} and quantum dots~\cite{Marcus1992}.  
In the context of single-particle quantum chaos, scars represent a concentration of some eigenfunctions along the trajectory of classical periodic orbits. In our study, \emph{many-body} scars form in the Hilbert space of an interacting quantum system, and they are identified with special eigenstates that can be effectively constructed by a tight-binding method. These special eigenstates, which occur at arbitrary energy densities,  are concentrated in parts of Hilbert space. Their experimental manifestation are the long-lived oscillations for certain (experimentally preparable) initial states~\cite{Bernien17}. 
The formation of quantum-scarred bands in many-body systems is facilitated by the lack of a simple tensor product structure of the constrained Hilbert space. Our results suggest that scarred many-body bands give rise to a new dynamical regime, which likely comprises a new universality class beyond conventional ergodic, integrable and MBL systems. This opens up opportunities for creating and manipulating novel states with long-lived quantum coherence in systems which are now amenable to experimental study. 

\emph{Model.---}In the experiment~\cite{Bernien17}, a chain of Rydberg atoms was realized in which, effectively, excitations were created/annihilated with equal amplitudes. In the limit where the nearest neighbor interaction is much larger than the detuning and the Rabi frequency, the system is modelled by the following spin-$1/2$ Hamiltonian~\cite{Lesanovsky2012}, 
\begin{equation}
  \label{Eq:Ham}
  H = \sum_{i=1}^L P_{i} X_{i+1} P_{i+2},
\end{equation}
where $X_i, Y_i, Z_i$ are the Pauli operators. With $|{\circ}{\rangle}$, $|{\bullet}{\rangle}$ being the two available states, the operator $X_i = |\circ\rangle\langle\bullet|+|\bullet\rangle\langle\circ|$ creates or removes an excitation at a given site, and projectors $ P_i=|\circ\rangle\langle\circ|=(1- Z_i)/{2}$, written in terms of $Z_i = |\bullet\rangle\langle\bullet|-|\circ\rangle\langle\circ|$, ensure that the nearby atoms are not simultaneously in the excited state. For example, $P_1 X_2 P_3$ acting on $|{\circ}{\circ}{\circ}\rangle$ gives $|{\circ}{\bullet}{\circ}\rangle$ (and vice versa), while it annihilates any of the configurations $|{\bullet}{\circ}{\circ}\rangle$, $|{\circ}{\circ}{\bullet}\rangle$, $|{\bullet}{\circ}{\bullet}\rangle$.

\begin{figure}[t]
\begin{center}
\includegraphics[width=0.9\columnwidth]{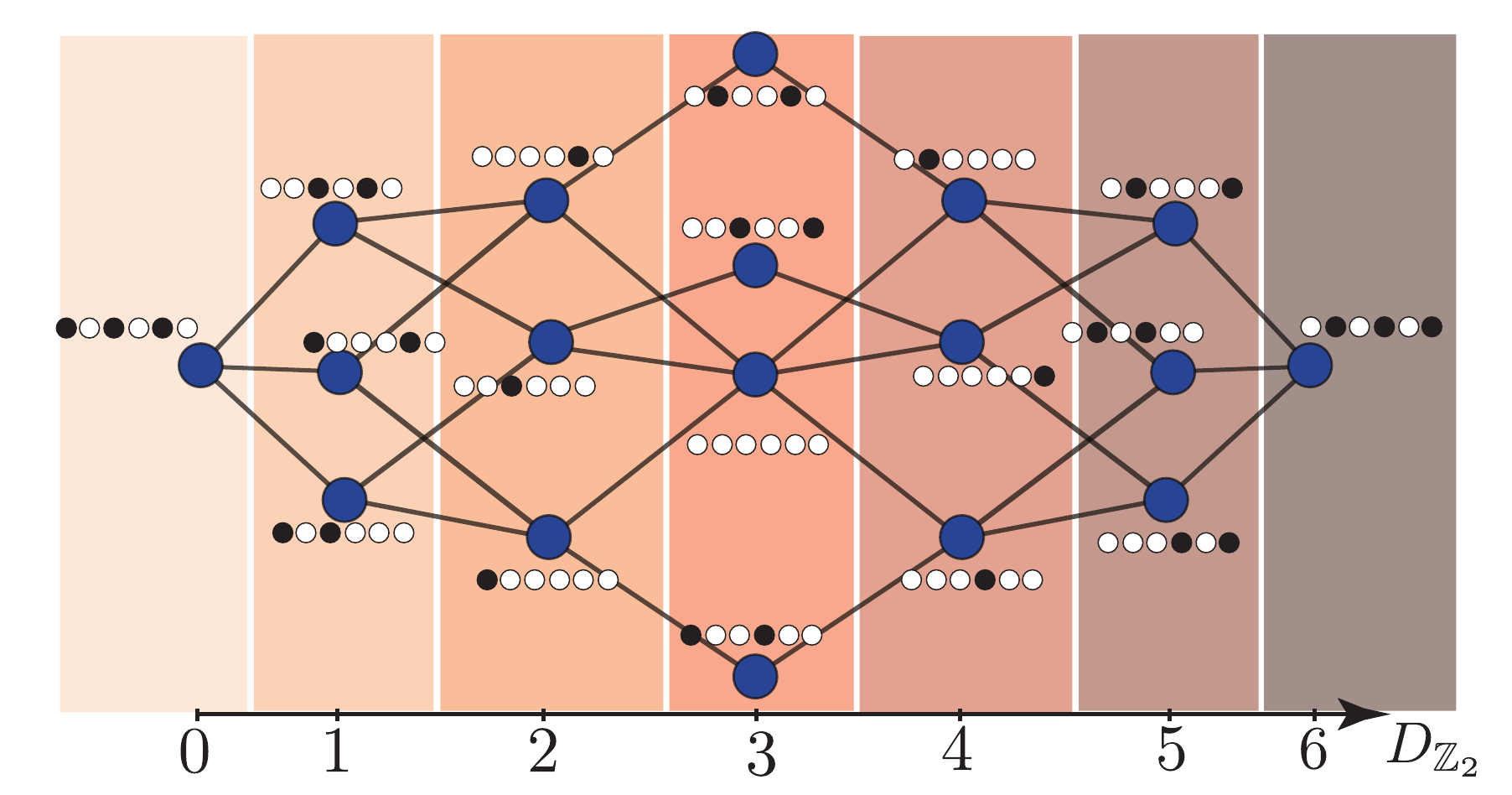}\\
\caption{ \label{Fig:model}
The action of the Hamiltonian (\ref{Eq:Ham}) on the Hilbert space of the Fibonacci chain for $L=6$ sites is represented as a graph where the nodes label the allowed product states, while the edges connect configurations that result from a given product state due to the action of Hamiltonian. Nodes of the graph are grouped according to the Hamming distance from $|\mathbb{Z}_2\rangle$ state, $D_{\mathbb{Z}_2}$, shown below.}
\end{center}
\end{figure}

Due to the presence of projectors in $H$, the Hilbert space of the model acquires a non-trivial local constraint. Namely, each atom can be either in the ground $|\circ\rangle$ or the excited state $|\bullet\rangle$, but excluding those configurations where two neighboring atoms are in the excited state, $|\ldots{\bullet}{\bullet}\ldots\rangle$. Such a constraint makes the Hilbert space identical to that of chains of non-Abelian Fibonacci anyons, rather than spins-$1/2$ or fermions.  
For periodic boundary conditions (PBC), the Hilbert space dimension is equal to ${\cal D}  = F_{L-1}+F_{L+1}$, where $F_n$ is the $n$th Fibonacci number. For instance, in the case of $L=6$ chain we have ${\cal D} = 18$, as shown in Fig.~\ref{Fig:model}.  For open boundary condition (OBC), ${\cal D}$ scales as $F_{L+2}$. Thus, the Hilbert space is evidently very different from, e.g., the spin-$\frac{1}{2}$ chain where the number of states grows as $2^L$. 

The model~(\ref{Eq:Ham}) is particle-hole symmetric: an operator ${\cal P}=\prod_i Z_i$ anticommutes with the Hamiltonian, $ {\cal P}H=-H {\cal P}$, and therefore each eigenstate $|\psi\rangle$ with energy $E\neq 0$ has a partner $ {\cal P}|\psi\rangle$ with energy $-E$. Furthermore, the model has spatial inversion symmetry $I$ which maps $i\to L-i+1$. In addition,  with PBC, this model has translation symmetry. In what follows, unless specified otherwise, we restrict to PBCs (thus identifying $i=L+1$ and~$i=1$) and explicitly resolve translation and inversion symmetries which allows us to fully diagonalize systems of up to $L=32$ sites (with ${\cal D}_{0+}=77436$ states in the zero momentum inversion-symmetric  sector). 

Experiment~\cite{Bernien17} and numerical simulations on small systems~\cite{Sun2008, LesanovskyDynamics} revealed that the relaxation under unitary dynamics specified by the Hamiltonian~(\ref{Eq:Ham}) strongly depends on the initial state of the system. In particular, starting from a period-2 CDW states 
\begin{equation}\label{Eq:Z2def}
|\mathbb{Z}_2\rangle =|{\bullet}{\circ}{\bullet}{\circ}\ldots\rangle,
\quad 
|\mathbb{Z}'_2\rangle =|{\circ}{\bullet}{\circ}{\bullet}\ldots\rangle,
\end{equation}
that are related by a translation by one lattice period, the system shows surprising long-time oscillations of local observables for long chains of up to $L=51$ sites. While this might suggest that the system is non-ergodic, it was also observed that the initial state with all atoms in the state $|\circ\rangle$ shows fast relaxation and no revivals, characteristic of thermalizing systems. Given that the model (\ref{Eq:Ham}) is translation invariant and has no disorder, MBL mechanism cannot be at play. Below we explain the origin of the observed oscillations and the apparent non-ergodic dynamics.

\begin{figure}[t]
  \includegraphics[width=0.95\columnwidth]{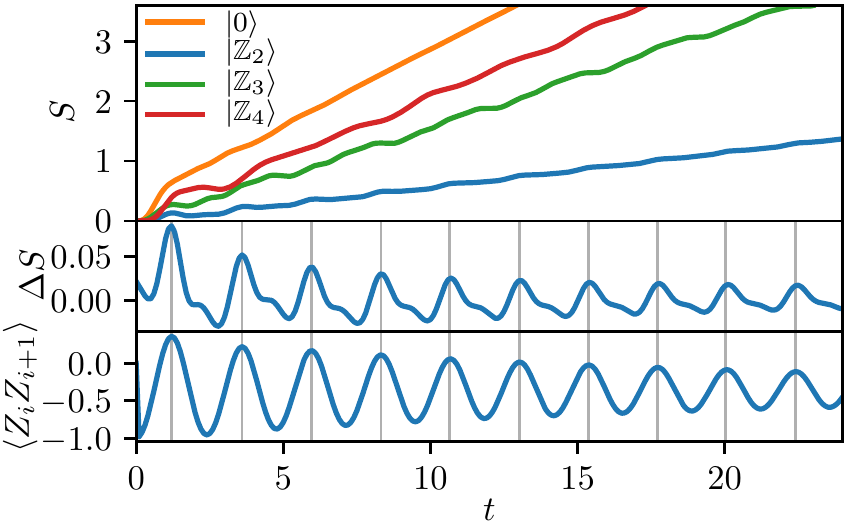}
  \caption{Entanglement entropy displays linear growth starting from various initial density-wave product states, as well as the fully polarized $|0\rangle=|\ldots{\circ}{\circ}{\circ}\ldots\rangle$ state.  Bottom panels illustrate that for $|\mathbb{Z}_2\rangle$ initial state entanglement oscillates around the linear growth with the same frequency as local correlation functions. \label{Fig:dynamics}}
\end{figure}

\begin{figure*}[t]
\includegraphics[width=\textwidth]{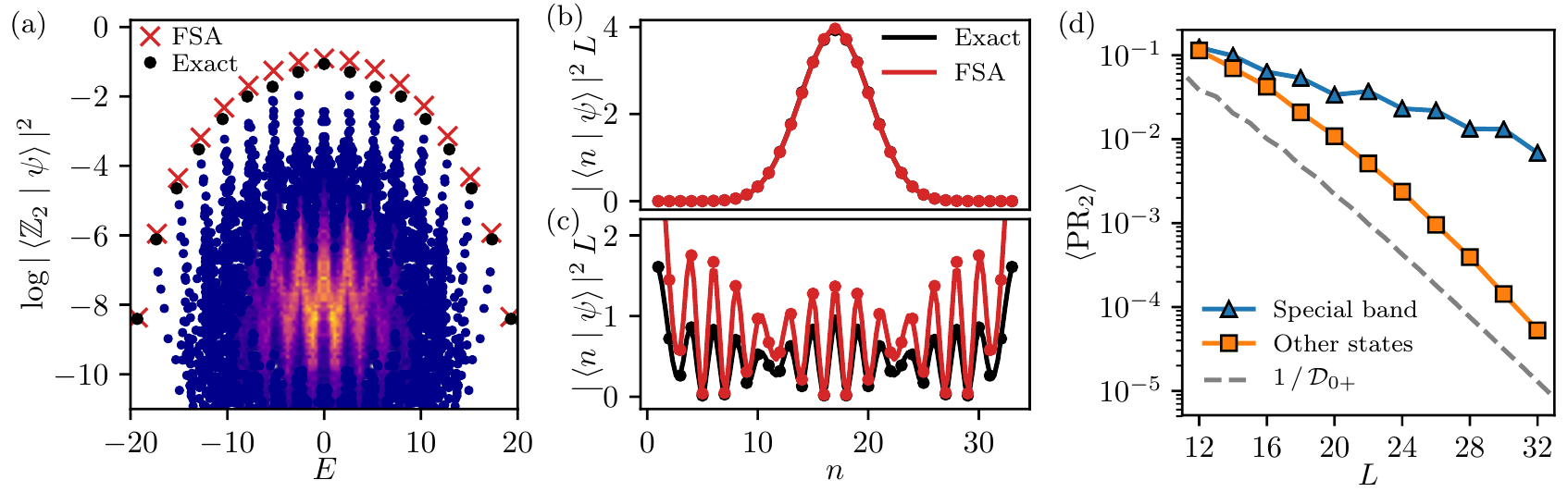}
  \caption{  \label{Fig:special}
    (a)~Scatter plot of the overlap of many-body eigenstates of the Hamiltonian~(\ref{Eq:Ham}) with $|\mathbb{Z}_2\rangle$ product state reveals a band of special eigenstates separated from the remaining eigenstates. Crosses denote overlaps with eigenstates from the FSA approximation, which agree very well with exact results. The density of data points (shown in the middle of the graph) illustrates the tower structure in the overlaps.
    (b-c)~Squared overlap between the basis vectors of FSA approximation $|n\rangle$ and exact eigenstates (black) or approximate FSA eigenstates (red).
     Panel (b) is for the ground state whereas the lower panel is for the state in the special band adjacent to energy $E=0$.
    (d)~Participation ratios of special eigenstates decay parametrically slower compared to the average participation ratio of all states within the same energy range. Dashed line shows the inverse Hilbert space dimension.
    All data is for $L=32$ in inversion-symmetric, zero momentum symmetry sector.
  }
\end{figure*}

 \emph{Dynamics.---}We start by characterizing dynamical evolution of the model~(\ref{Eq:Ham}) for different initial conditions. Motivated by experiment~\cite{Bernien17}, we consider a family of density-wave states $|\mathbb{Z}_k\rangle$ with period $k=2,3,4$, as well as the fully polarized state $|0\rangle$. We use infinite time evolving block decimation (iTEBD) method which provides results valid in  the thermodynamic limit up to some finite time~\cite{Vidal07}. The bond dimension used is 400, which limits the evolution time to about $t\sim 30$.
 
Top panel of Fig.~\ref{Fig:dynamics} reveals linear growth of entanglement entropy for all considered initial states. Yet, the slope of entanglement growth strongly depends on the initial state, with the slowest growth observed when the system is prepared in the period-2 density wave state,~$|\mathbb{Z}_2\rangle$, in Eq.~(\ref{Eq:Z2def}). 
In addition, the entanglement growth has weak oscillations on top of the linear growth, which are most pronounced for $|\mathbb{Z}_2\rangle$ initial state. Middle panel of Fig.~\ref{Fig:dynamics} illustrates the oscillations in entanglement by subtracting the linear component. We note that the oscillations are periodic with the period $T_{\mathbb{Z}_2}\approx 2.35$, in agreement with Ref.~\cite{Bernien17}. Similarly, periodic oscillations are clearly visible in the local correlation function, $\corr{Z_iZ_{i+1}}$.
The oscillations that persist for long times when entanglement light-cone reaches the distance of about $\gtrsim 20$ sites, as evidenced by the correlation function, are highly unusual. While experimental work~\cite{Bernien17} presented a variational ansatz capturing these oscillations, below we demonstrate that the oscillations actually arise due to the existence of special eigenstates within the rest of the many-body spectrum. 

\emph{Special states.---}The special eigenstates become clearly visible when one arranges the entire many-body spectrum according to the overlap with the density-wave $|\mathbb{Z}_2\rangle$ state, as shown in Fig.~\ref{Fig:special}(a). This reveals the ``$\mathbb{Z}_2$-band'' of  special eigenstates, which are distinguished by atypically high overlaps with the $|\mathbb{Z}_2\rangle$ product state. The energy separation between states stays approximately constant near the center of the band and equal to  $\Omega  \approx 1.33$. This energy separation matches half the frequency of the real-time oscillations observed in iTEBD numerical simulations in Fig.~\ref{Fig:dynamics}. The factor of 2 comes from the fact that the measured correlator does not distinguish between the  $|\mathbb{Z}_2\rangle$ and  $|\mathbb{Z}_2'\rangle$ states.

Next, we show that it is possible to construct accurate approximations to the entire band of special states. This is surprising because the model in Eq.~(\ref{Eq:Ham}) is not frustration free, hence even its ground state may not be expressible as MPS with bond dimension equal to 2~\cite{LesanovskyMPS}. Remarkably, the eigenstates in the $\mathbb{Z}_2$-band can still be accurately described within an effective tight-binding approximation. In this effective description, a ``site" will turn out to be a superposition of product states at fixed Hamming distance $D_{\mathbb{Z}_2}$ from the $|\mathbb{Z}_2\rangle$ product state, which is defined as the minimum number of spin flips required to transform those states into $| \mathbb{Z}_2\rangle$. 
 
 We start by splitting the  Hamiltonian as $H = H^{+} + H^{-}$, where we have introduced the operator 
\begin{eqnarray}\label{Eq:Hpm}
  \label{Eq:Hp}
    H^{+}  = \sum_{i\in\text{ even}} P_{i-1} \sigma^+_i P_{i+1} + \sum_{i \in\text{ odd}} P_{i-1} \sigma^-_i P_{i+1},
\end{eqnarray}
with $\sigma_i^+ = |\bullet\rangle\langle\circ|$ and $\sigma_i^- = |\circ\rangle\langle\bullet|$. $H^+$ increases $D_{\mathbb{Z}_2}$ by~1 (similarly, $H^-$ lowers $D_{\mathbb{Z}_2}$). In order to derive the tight-binding model, we follow the Lanczos procedure~\cite{Lanczos} for the Hamiltonian expressed in terms of Eq.~(\ref{Eq:Ham}) and the initial state $|\mathbb{Z}_2\rangle$. This will formally yield an effective system corresponding to a particle hopping on the lattice containing $L+1$ sites.

Lanczos algorithm is a transformation that reduces the Hamiltonian to a tridiagonal form. The basis vectors of the reduced form are iteratively calculated by acting with the Hamiltonian on a vector and then orthogonalizing against the previous basis vectors. The simplicity of  the Lanczos procedure is in the fact that it is sufficient to perform orthogonalization against only the last vector added to the basis. In the $j^{th}$ iteration of the algorithm, the new (unnormalized) vector in the basis is $|\tilde{u}_{j+1} \rangle = H |u_j \rangle  - \beta_{j-1} |u_{j-1} \rangle$, where $|u_{j-1} \rangle$ is the $(j-1)^{th}$ vector in the basis, $\beta_{j-1} = \langle u_{j-1} | H | u_j \rangle$ and tilde stands for the unnormalized vector. 

In our modified Lanczos procedure, we introduce the forward/backward propagation by $H^{\pm} |u_j \rangle$. If the backward propagation results to approximately the same vector as the previous one in the basis, i.e.,  $H^- |u_j \rangle \approx \beta_{j-1} |u_{j-1} \rangle$, this implies that $|\tilde{u}_{j+1} \rangle = H^+|u_j \rangle = \beta_{j}|u_{j+1} \rangle$.
In general, backward propagation would take us far from the vector $|u_{j-1}\rangle$, but in special cases like the Hamiltonian $\sum_i X_i$, it can result in the \emph{same} vector $|u_{j-1}\rangle$. To an excellent approximation, this is also true in our case with the added projectors $P_{i-1}$ and $P_{i+1}$, provided that the initial state is $|\mathbb{Z}_2\rangle$ state. In such cases, we can label the vectors $|u_n\rangle \equiv |n\rangle$ by their Hamming distance $n$ from the initial state. The major implication of our modified Lanczos procedure is that there is no intersection between the product state supports of different basis vectors $|n \rangle, |m \rangle$ for $m \neq n$, which implies that the algorithm must close after exactly $L$ iterations. 

To summarize, the Lanczos procedure we follow here can be  understood as a forward scattering approximation (FSA) on a lattice with sites labelled by the Hamming distance. The basis of the effective tight-binding model is $\{ |0\rangle,|1\rangle,\ldots,|L\rangle\}$, where $|0\rangle \equiv |\mathbb{Z}_2\rangle$ and $|n\rangle = (H^{+})^n|\mathbb{Z}_2\rangle/||(H^{+})^n|\mathbb{Z}_2\rangle||$. The tridiagonal matrix resulting from this procedure is the FSA Hamiltonian,
\begin{equation}\label{Eq:TB}
H_{FSA}= \sum_{n=0}^{L}\beta_{n}(|n \rangle \langle n+1| + h.c.),
\end{equation}
where the hopping amplitude is given by 
\begin{eqnarray}
\label{eg:beta}
\beta_{n}= \langle n+1| H^+|n \rangle = \langle n| H^-|n+1 \rangle.
\end{eqnarray}
This is an effective tight-binding model that captures the band of special states in Fig.~\ref{Fig:special}(a).

In the usual Lanczos procedure, there is no a priori reason for Eq.~(\ref{eg:beta}) to hold, and we can quantify the error per iteration of the FSA approximation by
$\text{err}(n) = |\langle n| H^+ H^- |n \rangle/\beta_{n}^2-1|$,
where $\text{err}(n) = 0$ is equivalent to  $H^- |n \rangle= \beta_{n-1} |n -1 \rangle$. Numerically we find that $\text{err}(n) \approx 0.2 \%$ for $L=32$ and has a decreasing trend as we increase the system size, which is promising in terms of scaling the method to the thermodynamic limit. As an additional error measure, the average energy difference between the exact eigenstates in the $\mathbb{Z}_2$-band and the eigenstates of $H_\text{FSA}$ for $L=32$ is $\overline{\Delta E/E}\approx 1\%$, which further supports the accuracy of this approximation scheme. Additional discussion of the errors and benchmarks of the method will be presented elsewhere.\footnoteremember{fn}{A. Michailidis \emph{et al.}, in preparation.}

Finally, we compare the eigenstates of $H_\text{FSA}$ with exact eigenstates from the special band obtained numerically in $L=32$ chain with PBC. In Fig.~\ref{Fig:special}(b)  we observe that  the lowest-energy special state has exactly the same overlaps with the basis states $\ket{n}$ as the FSA eigenstate. For the special eigenstates in the middle of many-body band, such as the one shown in Fig.~\ref{Fig:special}(c), the FSA overestimates the overlap, yet capturing the oscillations. The agreement between FSA and exact eigenstates is highly surprising, and it further supports the unusual nature of the special eigenstates. Indeed, a  basis that has only $L+1$ states, each concentrated in small parts of the Hilbert space, would provide an extremely poor approximation for a generic highly excited eigenstate of a thermalizing system of size $L$. 

In order to provide the further insights into the structure of special eigenstates, we study their participation ratios in the product state basis. The second participation ratio $\text{PR}_2$ of eigenstate $|\psi\rangle$ is defined as a sum of all wave function coefficients, $\text{PR}_2 = \sum_\alpha |\langle\alpha|\psi\rangle|^4 $, where $\alpha$ label all distinct product states in the inversion symmetric zero-momentum sector. For ergodic states, one expects that $\text{PR}_2$ decreases as inverse Hilbert space dimension of corresponding sector. This is indeed what we observe in Fig.~\ref{Fig:special}(d) for the $\langle \text{PR}_2\rangle_\text{av}$ averaged over all eigenstates (with $E\neq0$, see below) in the middle 2/3 of the full energy band. At the same time, $\text{PR}_2$ averaged over \emph{special} eigenstates from the same energy interval also decreases exponentially with the system size, yet being exponentially enhanced compared to  $\langle P_2\rangle_\text{av}$. The exponential enhancement of $\text{PR}_2$ is evident in Fig.~\ref{Fig:special}(d) and its persistence for chains of up to $L=32$ sites provides strong evidence for the existence of special states even in the thermodynamic limit.

In addition, the enhancement of the participation ratio for special eigenstates evidences their concentration in some subregions of the Hilbert space. This suggests that special eigenstates are equivalent to the many-body version of  quantum scarred wave functions which concentrate in vicinity of unstable periodic classical orbits in the single-particle quantum chaos. The success of FSA yielding a good approximation to special eigenstates shows that the ``special  trajectory'' in the present case connects the two different CDW states, $\ket{\mathbb{Z}_2}$ and $\ket{\mathbb{Z}'_2}$. This provides a natural explanation for the unusual dynamics observed in Fig~\ref{Fig:dynamics} and in experiment~\cite{Bernien17}, including the revivals in the many-body fidelity starting from $\ket{\mathbb{Z}_2}$ initial state. In addition, we have evidence that other states, e.g., $|{\bullet}{\circ}{\circ}{\bullet}{\circ}{\circ}\ldots\rangle$  lead to similar behavior to $|\mathbb{Z}_2\rangle$ state, although at present it is not clear how to determine all such configurations.

\begin{figure}[b]
\begin{center}
\includegraphics[width=0.9\columnwidth]{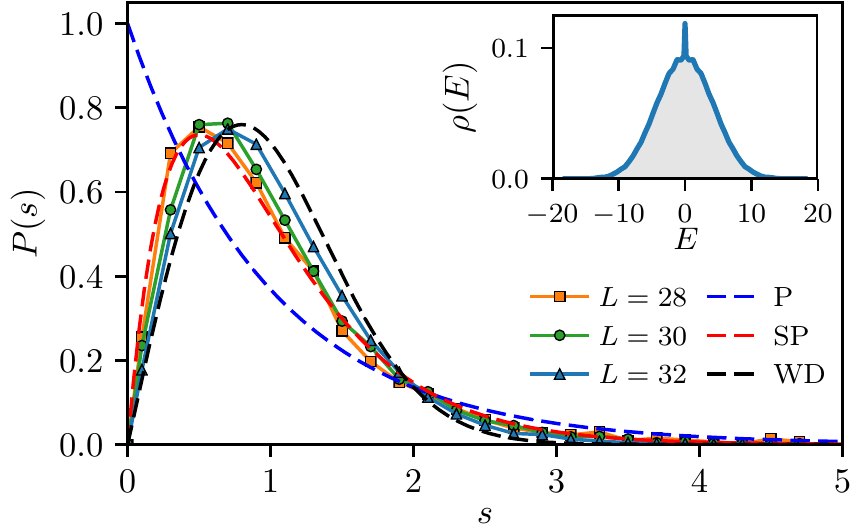}\\
\caption{ \label{Fig:levelstat}  The level statistics of $k=0$, $I=+1$ sector of the model in Eq.~(\ref{Eq:Ham}) interpolates between Semi-Poisson (SP) and Wigner-Dyson (WD) distributions with increasing system size.  Data is inconsistent with the Poisson statistics (P). The distribution is shown for unfolded energy levels $E_i$ with $i\in [{\cal D}_{0+}/5,{\cal D}_{0+}/2-500]$ to exclude the lowest part of the spectrum and the most central part, where zero modes are concentrated. The inset shows the density of states for $L=32$, illustrating that it has a Gaussian form, without any anomalies except for the spike at $E=0$ from zero modes.
}
\end{center}
\end{figure}

\emph{Absence of integrability.---} Above we suggested the connection between the special states, non-ergodic dynamics and quantum scars. However, an alternative explanation for the unusual behavior observed could be some proximate integrability~\cite{FendleySachdev, FendleyXYZ}. Indeed, the  Hamiltonian~(\ref{Eq:Ham}) can be seen as a deformation of the ``golden chain'' introduced by Feiguin \emph{et al.}~\cite{Feiguin07} which is Yang-Baxter integrable. However, the present model cannot be treated as a weak deformation of the ``golden chain''~\cite{Feiguin07}, as the two Hamiltonians differ by several terms with $O(1)$ coefficients. 

To investigate possible integrability, we studied the level statistics of the model in Eq.~(\ref{Eq:Ham}) in Fig.~\ref{Fig:levelstat}, which is a common diagnostics in the context of both single-particle and many-body quantum systems. Fig.~\ref{Fig:levelstat} reveals that even for relatively small system size, there is a pronounced level repulsion and the distribution of the energy level spacing is close to Semi-Poisson~\cite{Bogomolny99}. This is in sharp contrast with integrable systems which always have Poisson level statistics. Moreover, upon increasing the system size to $L=32$, we observe that the level statistics steadily approaches the Wigner-Dyson distribution. The Wigner-Dyson level statistics, along with ballistic growth of entanglement, rule out the integrability-based explanation of  the non-ergodic dynamics in the model~(\ref{Eq:Ham}). 

\emph{Zero modes.---}Another feature prominent in the inset of Fig.~\ref{Fig:levelstat} is the peak in the otherwise Gaussian density of states. This peak is caused by the
 \emph{exponentially large} number of degenerate eigenstates at $E=0$. For OBC, the number of zero modes scales as ${\cal Z}_L=F_{\frac{L}{2}+1}$ when $L$ is even, and ${\cal Z}_L=F_{\frac{L-1}{2}}$ when $L$ is odd. 
 
To derive the zero mode count and to understand their origin, we reformulate the problem of finding eigenstates as a hopping problem on a graph with vertices corresponding to product states in the constrained Hilbert space. An example of such a graph for $L=6$ was illustrated in Fig~\ref{Fig:model}. The quantum many-body problem becomes equivalent to a single-particle hopping on this graph. Since the application of any term in the Hamiltonian changes the number of excitations ($\bullet$) by $\pm1$, the graph has a bipartite structure with even/odd sublattices corresponding to the product states with an even/odd number of excitations. Equivalently, this follows from the existence of $\cal P$ which anticommutes with the Hamiltonian.

It is well-known that bipartite lattices support zero modes, with their number lower-bounded by the difference between the number of sites in two sublattices~\cite{Sutherland86,Inui}. In the present case the difference between sectors with an even and odd number of excitations is at most equal to one. However, one should take into account the inversion symmetry $ I$, which splits the Hilbert space into two sectors with $I=\pm 1$. Importantly, within each of the sectors, the bipartite structure is preserved, but the state counting changes. This is because the inversion-odd sector does not include any inversion-invariant product states, e.g. the basis state ${\circ}{\bullet}{\circ}{\circ}{\bullet}{\circ}$ does not belong to $I=-1$ sector of $L=6$ chain. When system size $L=2k$ is even, \emph{all} inversion-symmetric states have  \emph{even} number of excitations, and the number of such states is given by ${\cal Z}_{2k}=F_{k+1}$. Hence in $I=-1$ sector the even sublattice of the graph will have a deficit of inversion-symmetric states, leading to the presence of zero modes. In this way, we can directly classify the zero-mode states, whose number was stated above.  The counting of zero modes for PBC can be determined similarly.~\footnoterecall{fn} 

In addition, we note that zero modes are stable with respect to introducing potential energy as long as it commutes with particle-hole symmetry $\cal P$  and anticommutes with inversion operator. An example of such potential energy is provided by the staggered chemical potential, $\sum_{i=1}^L (-)^i Z_i$, which does not change the number of zero modes in the model (\ref{Eq:Ham}) with OBC for even system sizes. Incidentally, the addition of staggered field also preserves the graph structure of the model, hence the tight-binding approximation remains valid and we expect similar persistent oscillations when the system is quenched from a $|\mathbb{Z}_2\rangle$ state. This is indeed confirmed by iTEBD simulations for weak but finite staggered fields.~\footnoterecall{fn}

While our symmetry arguments allowed us to enumerate all zero modes, the complete understanding of their algebraic properties appears to be a much more difficult problem. Interestingly, the many-body wave functions of the zero mode eigenstates may be chosen such that they all have integer coefficients in the product state basis (modulo overall normalization). This might be anticipated from the fact that the Hamiltonian matrix elements are all integers, and the solutions to the zero mode condition $H|\psi\rangle=0$ can be obtained by Gaussian elimination. Nevertheless, the decomposition into integers is suggestive of the existence of a recursive relation which bears curious similarity to Jack polynomials that appear in the Calogero-Sutherland model~\cite{CalogeroSutherland} and the fractional quantum Hall effect~\cite{BernevigHaldane}. In the latter case, model Hamiltonians can be defined that enforce local constraints in direct analogy with Eq.~(\ref{Eq:Ham}), and the quasihole excitations (whose number grows exponentially, determined by the constrained Hilbert space) appear as exact zero modes of the Hamiltonian, with integer coefficients in the product state basis. While we do not believe there are direct physical similarities between these models, it would be interesting to elucidate their mathematical relations. 

\emph{Discussion.---}In summary, we have demonstrated the weak breakdown of eigenstate thermalization in the Fibonacci chain. This breakdown is associated with a band of special eigenstates that we identified as ``quantum many-body scars". These eigenstates are analogs of single-particle chaotic wave functions, but with scars concentrated in parts of the Hilbert space. Moreover, as shown transparently by our tight-binding method, the scars can be experimentally probed by initializing the system in special states, such as $|{\bullet}{\circ}{\bullet}{\circ}\ldots\rangle$ and $|{\circ}{\bullet}{\circ}{\bullet}\ldots\rangle$. The ensuing quantum dynamics then remains concentrated on a very specific subset of the Hilbert space, giving rise to robust oscillations even in very large systems. 

We emphasize that our model is qualitatively different from previously considered models where the possibility of ergodicity breaking was investigated~\cite{carleo2012localization, Huveneers13, Muller, QDL,
 Yao14, QDL2, QDLEssler, PhysRevB.90.165137, Kim2016,Yarloo2017,michailidis2017slow,Lan2017_2,Smith2017,brenes2017many}. These studies focused on translation-invariant models where potential energy was designed to make most of the processes off-resonant, similar to MBL. The model~(\ref{Eq:Ham}), in contrast, does not have any potential energy, and the non-ergodic behavior arises due to constraints. While similar  kinetically constrained models~\cite{KCM-rev} with flat potential energy landscape were demonstrated to have slow MBL-like dynamics~\cite{Juan15,Juan16} in certain regimes, the present model does not belong to this class due to the ballistic propagation of  entanglement.

Our study suggests the existence of a new universality class of quantum dynamics, which is neither fully thermalizing nor MBL, and which we attribute to the presence of a local dynamical constraint. This opens many exciting research directions which could lead to better understanding  of weakly non-thermalizing systems. In particular, the analogy with quantum scars should be put on firmer footing. This requires a more rigorous generalization of  the concept of ``trajectory" (in the sense of single-particle quantum scars) to the many-body case. Full classification and understanding of all such trajectories and their ``parent states'' also remains an open problem. Another open question concerns the precise relationship between quantum scars and other unusual aspects of the model, such as the existence of zero modes. The forward scattering approximation presented here may be regarded as a first step in these directions. Finally, extending the analogy to quantum scars, we also expect the observed non-ergodic behavior to be stable with respect to weak perturbations, the investigation of which we reserve for future work. 

While questions formulated above may be addressed in the context of the specific model~(\ref{Eq:Ham}),  our work motivates the search for similar behavior in different kinetically-constrained models. It would be highly desirable to understand the  features of constrained models which give rise to the non-ergodic dynamics. In particular,  model~(\ref{Eq:Ham}) is a ``projection'' of a trivial paramagnet Hamiltonian, thus the projection of other non-interacting models might be a promising direction in such a search. 

Provided one identifies a broader class of ``quantum scarred'' models, these can be used to engineer many-body states with long coherence times in future experiments.  Better understanding and construction of models with zero modes  can be also very useful, as such states would be exceptionally long-lived and it is feasible to approximately create them in the existing quantum simulators~\cite{Bloch2012, Blatt2012}. Moreover, their algebraic properties might be of practical interest for quantum computation. Such endeavours would hopefully also uncover the interesting connections between quantum dynamics and spectral theory of graphs, and lead to better understanding of different universality classes of thermalizing systems.

\emph{Acknowledgments.---}We acknowledge insigthful discussions with M. Lukin and Wen Wei Ho. 
C.T., A.M., and Z.P. acknowledge support by EPSRC grants EP/P009409/1 and EP/M50807X/1, and the Royal Society Research Grant RG160635. D.A. acknowledges support by the Swiss National Science Foundation. Statement of compliance with EPSRC policy framework on research data: This publication is theoretical work that does not require supporting research data. This work was initiated during ``Conference on Many-Body-Localization: Advances in the Theory and Experimental Progress" at ICTP Trieste.

\bibliography{qdyn}

\begin{thebibliography}{64}%
\makeatletter
\providecommand \@ifxundefined [1]{%
 \@ifx{#1\undefined}
}%
\providecommand \@ifnum [1]{%
 \ifnum #1\expandafter \@firstoftwo
 \else \expandafter \@secondoftwo
 \fi
}%
\providecommand \@ifx [1]{%
 \ifx #1\expandafter \@firstoftwo
 \else \expandafter \@secondoftwo
 \fi
}%
\providecommand \natexlab [1]{#1}%
\providecommand \enquote  [1]{``#1''}%
\providecommand \bibnamefont  [1]{#1}%
\providecommand \bibfnamefont [1]{#1}%
\providecommand \citenamefont [1]{#1}%
\providecommand \href@noop [0]{\@secondoftwo}%
\providecommand \href [0]{\begingroup \@sanitize@url \@href}%
\providecommand \@href[1]{\@@startlink{#1}\@@href}%
\providecommand \@@href[1]{\endgroup#1\@@endlink}%
\providecommand \@sanitize@url [0]{\catcode `\\12\catcode `\$12\catcode
  `\&12\catcode `\#12\catcode `\^12\catcode `\_12\catcode `\%12\relax}%
\providecommand \@@startlink[1]{}%
\providecommand \@@endlink[0]{}%
\providecommand \url  [0]{\begingroup\@sanitize@url \@url }%
\providecommand \@url [1]{\endgroup\@href {#1}{\urlprefix }}%
\providecommand \urlprefix  [0]{URL }%
\providecommand \Eprint [0]{\href }%
\providecommand \doibase [0]{http://dx.doi.org/}%
\providecommand \selectlanguage [0]{\@gobble}%
\providecommand \bibinfo  [0]{\@secondoftwo}%
\providecommand \bibfield  [0]{\@secondoftwo}%
\providecommand \translation [1]{[#1]}%
\providecommand \BibitemOpen [0]{}%
\providecommand \bibitemStop [0]{}%
\providecommand \bibitemNoStop [0]{.\EOS\space}%
\providecommand \EOS [0]{\spacefactor3000\relax}%
\providecommand \BibitemShut  [1]{\csname bibitem#1\endcsname}%
\let\auto@bib@innerbib\@empty
%</preamble>
\bibitem [{\citenamefont {Kinoshita}\ \emph {et~al.}(2006)\citenamefont
  {Kinoshita}, \citenamefont {Wenger},\ and\ \citenamefont
  {Weiss}}]{Kinoshita06}%
  \BibitemOpen
  \bibfield  {author} {\bibinfo {author} {\bibfnamefont {Toshiya}\ \bibnamefont
  {Kinoshita}}, \bibinfo {author} {\bibfnamefont {Trevor}\ \bibnamefont
  {Wenger}}, \ and\ \bibinfo {author} {\bibfnamefont {David~S.}\ \bibnamefont
  {Weiss}},\ }\bibfield  {title} {\enquote {\bibinfo {title} {A quantum
  newton's cradle},}\ }\href {\doibase 10.1038/nature04693} {\bibfield
  {journal} {\bibinfo  {journal} {Nature}\ }\textbf {\bibinfo {volume} {440}},\
  \bibinfo {pages} {900--903} (\bibinfo {year} {2006})}\BibitemShut {NoStop}%
\bibitem [{\citenamefont {Schreiber}\ \emph {et~al.}(2015)\citenamefont
  {Schreiber}, \citenamefont {Hodgman}, \citenamefont {Bordia}, \citenamefont
  {L{\"u}schen}, \citenamefont {Fischer}, \citenamefont {Vosk}, \citenamefont
  {Altman}, \citenamefont {Schneider},\ and\ \citenamefont {Bloch}}]{Bloch15}%
  \BibitemOpen
  \bibfield  {author} {\bibinfo {author} {\bibfnamefont {Michael}\ \bibnamefont
  {Schreiber}}, \bibinfo {author} {\bibfnamefont {Sean~S.}\ \bibnamefont
  {Hodgman}}, \bibinfo {author} {\bibfnamefont {Pranjal}\ \bibnamefont
  {Bordia}}, \bibinfo {author} {\bibfnamefont {Henrik~P.}\ \bibnamefont
  {L{\"u}schen}}, \bibinfo {author} {\bibfnamefont {Mark~H.}\ \bibnamefont
  {Fischer}}, \bibinfo {author} {\bibfnamefont {Ronen}\ \bibnamefont {Vosk}},
  \bibinfo {author} {\bibfnamefont {Ehud}\ \bibnamefont {Altman}}, \bibinfo
  {author} {\bibfnamefont {Ulrich}\ \bibnamefont {Schneider}}, \ and\ \bibinfo
  {author} {\bibfnamefont {Immanuel}\ \bibnamefont {Bloch}},\ }\bibfield
  {title} {\enquote {\bibinfo {title} {Observation of many-body localization of
  interacting fermions in a quasirandom optical lattice},}\ }\href {\doibase
  10.1126/science.aaa7432} {\bibfield  {journal} {\bibinfo  {journal}
  {Science}\ }\textbf {\bibinfo {volume} {349}},\ \bibinfo {pages} {842--845}
  (\bibinfo {year} {2015})},\ \Eprint
  {http://arxiv.org/abs/http://science.sciencemag.org/content/349/6250/842.full.pdf}
  {http://science.sciencemag.org/content/349/6250/842.full.pdf} \BibitemShut
  {NoStop}%
\bibitem [{\citenamefont {Smith}\ \emph {et~al.}(2016)\citenamefont {Smith},
  \citenamefont {Lee}, \citenamefont {Richerme}, \citenamefont {Neyenhuis},
  \citenamefont {Hess}, \citenamefont {Hauke}, \citenamefont {Heyl},
  \citenamefont {Huse},\ and\ \citenamefont {Monroe}}]{Monroe16}%
  \BibitemOpen
  \bibfield  {author} {\bibinfo {author} {\bibfnamefont {J.}~\bibnamefont
  {Smith}}, \bibinfo {author} {\bibfnamefont {A.}~\bibnamefont {Lee}}, \bibinfo
  {author} {\bibfnamefont {P.}~\bibnamefont {Richerme}}, \bibinfo {author}
  {\bibfnamefont {B.}~\bibnamefont {Neyenhuis}}, \bibinfo {author}
  {\bibfnamefont {P.~W.}\ \bibnamefont {Hess}}, \bibinfo {author}
  {\bibfnamefont {P.}~\bibnamefont {Hauke}}, \bibinfo {author} {\bibfnamefont
  {M.}~\bibnamefont {Heyl}}, \bibinfo {author} {\bibfnamefont {D.~A.}\
  \bibnamefont {Huse}}, \ and\ \bibinfo {author} {\bibfnamefont
  {C.}~\bibnamefont {Monroe}},\ }\bibfield  {title} {\enquote {\bibinfo {title}
  {Many-body localization in a quantum simulator with programmable random
  disorder},}\ }\href {http://dx.doi.org/10.1038/nphys3783} {\bibfield
  {journal} {\bibinfo  {journal} {Nat Phys}\ }\textbf {\bibinfo {volume}
  {12}},\ \bibinfo {pages} {907--911} (\bibinfo {year} {2016})}\BibitemShut
  {NoStop}%
\bibitem [{\citenamefont {{Kucsko}}\ \emph {et~al.}(2016)\citenamefont
  {{Kucsko}}, \citenamefont {{Choi}}, \citenamefont {{Choi}}, \citenamefont
  {{Maurer}}, \citenamefont {{Sumiya}}, \citenamefont {{Onoda}}, \citenamefont
  {{Isoya}}, \citenamefont {{Jelezko}}, \citenamefont {{Demler}}, \citenamefont
  {{Yao}},\ and\ \citenamefont {{Lukin}}}]{Lukin16}%
  \BibitemOpen
  \bibfield  {author} {\bibinfo {author} {\bibfnamefont {G.}~\bibnamefont
  {{Kucsko}}}, \bibinfo {author} {\bibfnamefont {S.}~\bibnamefont {{Choi}}},
  \bibinfo {author} {\bibfnamefont {J.}~\bibnamefont {{Choi}}}, \bibinfo
  {author} {\bibfnamefont {P.~C.}\ \bibnamefont {{Maurer}}}, \bibinfo {author}
  {\bibfnamefont {H.}~\bibnamefont {{Sumiya}}}, \bibinfo {author}
  {\bibfnamefont {S.}~\bibnamefont {{Onoda}}}, \bibinfo {author} {\bibfnamefont
  {J.}~\bibnamefont {{Isoya}}}, \bibinfo {author} {\bibfnamefont
  {F.}~\bibnamefont {{Jelezko}}}, \bibinfo {author} {\bibfnamefont
  {E.}~\bibnamefont {{Demler}}}, \bibinfo {author} {\bibfnamefont {N.~Y.}\
  \bibnamefont {{Yao}}}, \ and\ \bibinfo {author} {\bibfnamefont {M.~D.}\
  \bibnamefont {{Lukin}}},\ }\bibfield  {title} {\enquote {\bibinfo {title}
  {{Critical thermalization of a disordered dipolar spin system in diamond}},}\
  }\href@noop {} {\bibfield  {journal} {\bibinfo  {journal} {ArXiv e-prints}\ }
  (\bibinfo {year} {2016})},\ \Eprint {http://arxiv.org/abs/1609.08216}
  {arXiv:1609.08216 [cond-mat.mes-hall]} \BibitemShut {NoStop}%
\bibitem [{\citenamefont {Sutherland}(2004)}]{Sutherland}%
  \BibitemOpen
  \bibfield  {author} {\bibinfo {author} {\bibfnamefont {B.}~\bibnamefont
  {Sutherland}},\ }\href {http://books.google.com/books?id=aVUdnwEACAAJ} {\emph
  {\bibinfo {title} {Beautiful Models: 70 Years of Exactly Solved Quantum
  Many-body Problems}}}\ (\bibinfo  {publisher} {World Scientific},\ \bibinfo
  {year} {2004})\BibitemShut {NoStop}%
\bibitem [{\citenamefont {Basko}\ \emph {et~al.}(2006)\citenamefont {Basko},
  \citenamefont {Aleiner},\ and\ \citenamefont {Altshuler}}]{Basko06}%
  \BibitemOpen
  \bibfield  {author} {\bibinfo {author} {\bibfnamefont {D.M.}\ \bibnamefont
  {Basko}}, \bibinfo {author} {\bibfnamefont {I.L.}\ \bibnamefont {Aleiner}}, \
  and\ \bibinfo {author} {\bibfnamefont {B.L.}\ \bibnamefont {Altshuler}},\
  }\bibfield  {title} {\enquote {\bibinfo {title} {Metal--insulator transition
  in a weakly interacting many-electron system with localized single-particle
  states},}\ }\href {\doibase http://dx.doi.org/10.1016/j.aop.2005.11.014}
  {\bibfield  {journal} {\bibinfo  {journal} {Annals of Physics}\ }\textbf
  {\bibinfo {volume} {321}},\ \bibinfo {pages} {1126 -- 1205} (\bibinfo {year}
  {2006})}\BibitemShut {NoStop}%
\bibitem [{\citenamefont {Serbyn}\ \emph {et~al.}(2013)\citenamefont {Serbyn},
  \citenamefont {Papi\ifmmode~\acute{c}\else \'{c}\fi{}},\ and\ \citenamefont
  {Abanin}}]{Serbyn13-1}%
  \BibitemOpen
  \bibfield  {author} {\bibinfo {author} {\bibfnamefont {Maksym}\ \bibnamefont
  {Serbyn}}, \bibinfo {author} {\bibfnamefont {Z.}~\bibnamefont
  {Papi\ifmmode~\acute{c}\else \'{c}\fi{}}}, \ and\ \bibinfo {author}
  {\bibfnamefont {Dmitry~A.}\ \bibnamefont {Abanin}},\ }\bibfield  {title}
  {\enquote {\bibinfo {title} {Local conservation laws and the structure of the
  many-body localized states},}\ }\href {\doibase
  10.1103/PhysRevLett.111.127201} {\bibfield  {journal} {\bibinfo  {journal}
  {Phys. Rev. Lett.}\ }\textbf {\bibinfo {volume} {111}},\ \bibinfo {pages}
  {127201} (\bibinfo {year} {2013})}\BibitemShut {NoStop}%
\bibitem [{\citenamefont {Huse}\ \emph {et~al.}(2014)\citenamefont {Huse},
  \citenamefont {Nandkishore},\ and\ \citenamefont {Oganesyan}}]{Huse13}%
  \BibitemOpen
  \bibfield  {author} {\bibinfo {author} {\bibfnamefont {David~A.}\
  \bibnamefont {Huse}}, \bibinfo {author} {\bibfnamefont {Rahul}\ \bibnamefont
  {Nandkishore}}, \ and\ \bibinfo {author} {\bibfnamefont {Vadim}\ \bibnamefont
  {Oganesyan}},\ }\bibfield  {title} {\enquote {\bibinfo {title} {Phenomenology
  of fully many-body-localized systems},}\ }\href {\doibase
  10.1103/PhysRevB.90.174202} {\bibfield  {journal} {\bibinfo  {journal} {Phys.
  Rev. B}\ }\textbf {\bibinfo {volume} {90}},\ \bibinfo {pages} {174202}
  (\bibinfo {year} {2014})}\BibitemShut {NoStop}%
\bibitem [{\citenamefont {Deutsch}(1991)}]{DeutschETH}%
  \BibitemOpen
  \bibfield  {author} {\bibinfo {author} {\bibfnamefont {J.~M.}\ \bibnamefont
  {Deutsch}},\ }\href@noop {} {\bibfield  {journal} {\bibinfo  {journal} {Phys.
  Rev. A}\ }\textbf {\bibinfo {volume} {43}},\ \bibinfo {pages} {2146}
  (\bibinfo {year} {1991})}\BibitemShut {NoStop}%
\bibitem [{\citenamefont {Srednicki}(1994)}]{SrednickiETH}%
  \BibitemOpen
  \bibfield  {author} {\bibinfo {author} {\bibfnamefont {Mark}\ \bibnamefont
  {Srednicki}},\ }\bibfield  {title} {\enquote {\bibinfo {title} {Chaos and
  quantum thermalization},}\ }\href {\doibase 10.1103/PhysRevE.50.888}
  {\bibfield  {journal} {\bibinfo  {journal} {Phys. Rev. E}\ }\textbf {\bibinfo
  {volume} {50}},\ \bibinfo {pages} {888--901} (\bibinfo {year}
  {1994})}\BibitemShut {NoStop}%
\bibitem [{\citenamefont {Berry}\ and\ \citenamefont
  {Tabor}(1977)}]{BerryTabor}%
  \BibitemOpen
  \bibfield  {author} {\bibinfo {author} {\bibfnamefont {M.~V.}\ \bibnamefont
  {Berry}}\ and\ \bibinfo {author} {\bibfnamefont {M.}~\bibnamefont {Tabor}},\
  }\bibfield  {title} {\enquote {\bibinfo {title} {Level clustering in the
  regular spectrum},}\ }\href {\doibase 10.1098/rspa.1977.0140} {\bibfield
  {journal} {\bibinfo  {journal} {Proceedings of the Royal Society of London A:
  Mathematical, Physical and Engineering Sciences}\ }\textbf {\bibinfo {volume}
  {356}},\ \bibinfo {pages} {375--394} (\bibinfo {year} {1977})}\BibitemShut
  {NoStop}%
\bibitem [{\citenamefont {Pal}\ and\ \citenamefont {Huse}(2010)}]{PalHuse}%
  \BibitemOpen
  \bibfield  {author} {\bibinfo {author} {\bibfnamefont {Arijeet}\ \bibnamefont
  {Pal}}\ and\ \bibinfo {author} {\bibfnamefont {David~A.}\ \bibnamefont
  {Huse}},\ }\bibfield  {title} {\enquote {\bibinfo {title} {Many-body
  localization phase transition},}\ }\href {\doibase
  10.1103/PhysRevB.82.174411} {\bibfield  {journal} {\bibinfo  {journal} {Phys.
  Rev. B}\ }\textbf {\bibinfo {volume} {82}},\ \bibinfo {pages} {174411}
  (\bibinfo {year} {2010})}\BibitemShut {NoStop}%
\bibitem [{\citenamefont {Rigol}\ \emph {et~al.}(2007)\citenamefont {Rigol},
  \citenamefont {Dunjko}, \citenamefont {Yurovsky},\ and\ \citenamefont
  {Olshanii}}]{Rigol07}%
  \BibitemOpen
  \bibfield  {author} {\bibinfo {author} {\bibfnamefont {Marcos}\ \bibnamefont
  {Rigol}}, \bibinfo {author} {\bibfnamefont {Vanja}\ \bibnamefont {Dunjko}},
  \bibinfo {author} {\bibfnamefont {Vladimir}\ \bibnamefont {Yurovsky}}, \ and\
  \bibinfo {author} {\bibfnamefont {Maxim}\ \bibnamefont {Olshanii}},\
  }\bibfield  {title} {\enquote {\bibinfo {title} {Relaxation in a completely
  integrable many-body quantum system: An \textit{Ab~Initio} study of the
  dynamics of the highly excited states of 1d lattice hard-core bosons},}\
  }\href {\doibase 10.1103/PhysRevLett.98.050405} {\bibfield  {journal}
  {\bibinfo  {journal} {Phys. Rev. Lett.}\ }\textbf {\bibinfo {volume} {98}},\
  \bibinfo {pages} {050405} (\bibinfo {year} {2007})}\BibitemShut {NoStop}%
\bibitem [{\citenamefont {Rigol}\ \emph {et~al.}(2008)\citenamefont {Rigol},
  \citenamefont {Dunjko},\ and\ \citenamefont {Olshanii}}]{RigolNature}%
  \BibitemOpen
  \bibfield  {author} {\bibinfo {author} {\bibfnamefont {Marcos}\ \bibnamefont
  {Rigol}}, \bibinfo {author} {\bibfnamefont {Vanja}\ \bibnamefont {Dunjko}}, \
  and\ \bibinfo {author} {\bibfnamefont {Maxim}\ \bibnamefont {Olshanii}},\
  }\bibfield  {title} {\enquote {\bibinfo {title} {Thermalization and its
  mechanism for generic isolated quantum systems},}\ }\href {\doibase
  10.1038/nature06838} {\bibfield  {journal} {\bibinfo  {journal} {Nature}\
  }\textbf {\bibinfo {volume} {452}},\ \bibinfo {pages} {854--858} (\bibinfo
  {year} {2008})}\BibitemShut {NoStop}%
\bibitem [{\citenamefont {Kim}\ \emph {et~al.}(2014)\citenamefont {Kim},
  \citenamefont {Ikeda},\ and\ \citenamefont {Huse}}]{Huse14}%
  \BibitemOpen
  \bibfield  {author} {\bibinfo {author} {\bibfnamefont {Hyungwon}\
  \bibnamefont {Kim}}, \bibinfo {author} {\bibfnamefont {Tatsuhiko~N.}\
  \bibnamefont {Ikeda}}, \ and\ \bibinfo {author} {\bibfnamefont {David~A.}\
  \bibnamefont {Huse}},\ }\bibfield  {title} {\enquote {\bibinfo {title}
  {Testing whether all eigenstates obey the eigenstate thermalization
  hypothesis},}\ }\href {\doibase 10.1103/PhysRevE.90.052105} {\bibfield
  {journal} {\bibinfo  {journal} {Phys. Rev. E}\ }\textbf {\bibinfo {volume}
  {90}},\ \bibinfo {pages} {052105} (\bibinfo {year} {2014})}\BibitemShut
  {NoStop}%
\bibitem [{\citenamefont {Feiguin}\ \emph {et~al.}(2007)\citenamefont
  {Feiguin}, \citenamefont {Trebst}, \citenamefont {Ludwig}, \citenamefont
  {Troyer}, \citenamefont {Kitaev}, \citenamefont {Wang},\ and\ \citenamefont
  {Freedman}}]{Feiguin07}%
  \BibitemOpen
  \bibfield  {author} {\bibinfo {author} {\bibfnamefont {Adrian}\ \bibnamefont
  {Feiguin}}, \bibinfo {author} {\bibfnamefont {Simon}\ \bibnamefont {Trebst}},
  \bibinfo {author} {\bibfnamefont {Andreas W.~W.}\ \bibnamefont {Ludwig}},
  \bibinfo {author} {\bibfnamefont {Matthias}\ \bibnamefont {Troyer}}, \bibinfo
  {author} {\bibfnamefont {Alexei}\ \bibnamefont {Kitaev}}, \bibinfo {author}
  {\bibfnamefont {Zhenghan}\ \bibnamefont {Wang}}, \ and\ \bibinfo {author}
  {\bibfnamefont {Michael~H.}\ \bibnamefont {Freedman}},\ }\bibfield  {title}
  {\enquote {\bibinfo {title} {Interacting anyons in topological quantum
  liquids: The golden chain},}\ }\href {\doibase 10.1103/PhysRevLett.98.160409}
  {\bibfield  {journal} {\bibinfo  {journal} {Phys. Rev. Lett.}\ }\textbf
  {\bibinfo {volume} {98}},\ \bibinfo {pages} {160409} (\bibinfo {year}
  {2007})}\BibitemShut {NoStop}%
\bibitem [{\citenamefont {Trebst}\ \emph {et~al.}(2008)\citenamefont {Trebst},
  \citenamefont {Ardonne}, \citenamefont {Feiguin}, \citenamefont {Huse},
  \citenamefont {Ludwig},\ and\ \citenamefont {Troyer}}]{Trebst2008}%
  \BibitemOpen
  \bibfield  {author} {\bibinfo {author} {\bibfnamefont {Simon}\ \bibnamefont
  {Trebst}}, \bibinfo {author} {\bibfnamefont {Eddy}\ \bibnamefont {Ardonne}},
  \bibinfo {author} {\bibfnamefont {Adrian}\ \bibnamefont {Feiguin}}, \bibinfo
  {author} {\bibfnamefont {David~A.}\ \bibnamefont {Huse}}, \bibinfo {author}
  {\bibfnamefont {Andreas W.~W.}\ \bibnamefont {Ludwig}}, \ and\ \bibinfo
  {author} {\bibfnamefont {Matthias}\ \bibnamefont {Troyer}},\ }\bibfield
  {title} {\enquote {\bibinfo {title} {Collective states of interacting
  fibonacci anyons},}\ }\href {\doibase 10.1103/PhysRevLett.101.050401}
  {\bibfield  {journal} {\bibinfo  {journal} {Phys. Rev. Lett.}\ }\textbf
  {\bibinfo {volume} {101}},\ \bibinfo {pages} {050401} (\bibinfo {year}
  {2008})}\BibitemShut {NoStop}%
\bibitem [{\citenamefont {Lesanovsky}\ and\ \citenamefont
  {Katsura}(2012)}]{Lesanovsky2012}%
  \BibitemOpen
  \bibfield  {author} {\bibinfo {author} {\bibfnamefont {Igor}\ \bibnamefont
  {Lesanovsky}}\ and\ \bibinfo {author} {\bibfnamefont {Hosho}\ \bibnamefont
  {Katsura}},\ }\bibfield  {title} {\enquote {\bibinfo {title} {Interacting
  fibonacci anyons in a rydberg gas},}\ }\href {\doibase
  10.1103/PhysRevA.86.041601} {\bibfield  {journal} {\bibinfo  {journal} {Phys.
  Rev. A}\ }\textbf {\bibinfo {volume} {86}},\ \bibinfo {pages} {041601}
  (\bibinfo {year} {2012})}\BibitemShut {NoStop}%
\bibitem [{\citenamefont {Lindner}\ \emph {et~al.}(2012)\citenamefont
  {Lindner}, \citenamefont {Berg}, \citenamefont {Refael},\ and\ \citenamefont
  {Stern}}]{Lindner2012}%
  \BibitemOpen
  \bibfield  {author} {\bibinfo {author} {\bibfnamefont {Netanel~H.}\
  \bibnamefont {Lindner}}, \bibinfo {author} {\bibfnamefont {Erez}\
  \bibnamefont {Berg}}, \bibinfo {author} {\bibfnamefont {Gil}\ \bibnamefont
  {Refael}}, \ and\ \bibinfo {author} {\bibfnamefont {Ady}\ \bibnamefont
  {Stern}},\ }\bibfield  {title} {\enquote {\bibinfo {title} {Fractionalizing
  majorana fermions: Non-abelian statistics on the edges of abelian quantum
  hall states},}\ }\href {\doibase 10.1103/PhysRevX.2.041002} {\bibfield
  {journal} {\bibinfo  {journal} {Phys. Rev. X}\ }\textbf {\bibinfo {volume}
  {2}},\ \bibinfo {pages} {041002} (\bibinfo {year} {2012})}\BibitemShut
  {NoStop}%
\bibitem [{\citenamefont {Glaetzle}\ \emph {et~al.}(2014)\citenamefont
  {Glaetzle}, \citenamefont {Dalmonte}, \citenamefont {Nath}, \citenamefont
  {Rousochatzakis}, \citenamefont {Moessner},\ and\ \citenamefont
  {Zoller}}]{Glaetzle2014}%
  \BibitemOpen
  \bibfield  {author} {\bibinfo {author} {\bibfnamefont {A.~W.}\ \bibnamefont
  {Glaetzle}}, \bibinfo {author} {\bibfnamefont {M.}~\bibnamefont {Dalmonte}},
  \bibinfo {author} {\bibfnamefont {R.}~\bibnamefont {Nath}}, \bibinfo {author}
  {\bibfnamefont {I.}~\bibnamefont {Rousochatzakis}}, \bibinfo {author}
  {\bibfnamefont {R.}~\bibnamefont {Moessner}}, \ and\ \bibinfo {author}
  {\bibfnamefont {P.}~\bibnamefont {Zoller}},\ }\bibfield  {title} {\enquote
  {\bibinfo {title} {Quantum spin-ice and dimer models with rydberg atoms},}\
  }\href {\doibase 10.1103/PhysRevX.4.041037} {\bibfield  {journal} {\bibinfo
  {journal} {Phys. Rev. X}\ }\textbf {\bibinfo {volume} {4}},\ \bibinfo {pages}
  {041037} (\bibinfo {year} {2014})}\BibitemShut {NoStop}%
\bibitem [{\citenamefont {Vasseur}\ \emph {et~al.}(2015)\citenamefont
  {Vasseur}, \citenamefont {Potter},\ and\ \citenamefont
  {Parameswaran}}]{Vasseur2015}%
  \BibitemOpen
  \bibfield  {author} {\bibinfo {author} {\bibfnamefont {R.}~\bibnamefont
  {Vasseur}}, \bibinfo {author} {\bibfnamefont {A.~C.}\ \bibnamefont {Potter}},
  \ and\ \bibinfo {author} {\bibfnamefont {S.~A.}\ \bibnamefont
  {Parameswaran}},\ }\bibfield  {title} {\enquote {\bibinfo {title} {Quantum
  criticality of hot random spin chains},}\ }\href {\doibase
  10.1103/PhysRevLett.114.217201} {\bibfield  {journal} {\bibinfo  {journal}
  {Phys. Rev. Lett.}\ }\textbf {\bibinfo {volume} {114}},\ \bibinfo {pages}
  {217201} (\bibinfo {year} {2015})}\BibitemShut {NoStop}%
\bibitem [{\citenamefont {Chandran}\ \emph {et~al.}(2016)\citenamefont
  {Chandran}, \citenamefont {Schulz},\ and\ \citenamefont
  {Burnell}}]{Chandran16}%
  \BibitemOpen
  \bibfield  {author} {\bibinfo {author} {\bibfnamefont {A.}~\bibnamefont
  {Chandran}}, \bibinfo {author} {\bibfnamefont {Marc~D.}\ \bibnamefont
  {Schulz}}, \ and\ \bibinfo {author} {\bibfnamefont {F.~J.}\ \bibnamefont
  {Burnell}},\ }\bibfield  {title} {\enquote {\bibinfo {title} {The eigenstate
  thermalization hypothesis in constrained hilbert spaces: A case study in
  non-abelian anyon chains},}\ }\href {\doibase 10.1103/PhysRevB.94.235122}
  {\bibfield  {journal} {\bibinfo  {journal} {Phys. Rev. B}\ }\textbf {\bibinfo
  {volume} {94}},\ \bibinfo {pages} {235122} (\bibinfo {year}
  {2016})}\BibitemShut {NoStop}%
\bibitem [{\citenamefont {Lan}\ and\ \citenamefont {Powell}(2017)}]{Lan2017}%
  \BibitemOpen
  \bibfield  {author} {\bibinfo {author} {\bibfnamefont {Zhihao}\ \bibnamefont
  {Lan}}\ and\ \bibinfo {author} {\bibfnamefont {Stephen}\ \bibnamefont
  {Powell}},\ }\bibfield  {title} {\enquote {\bibinfo {title} {Eigenstate
  thermalization hypothesis in quantum dimer models},}\ }\href {\doibase
  10.1103/PhysRevB.96.115140} {\bibfield  {journal} {\bibinfo  {journal} {Phys.
  Rev. B}\ }\textbf {\bibinfo {volume} {96}},\ \bibinfo {pages} {115140}
  (\bibinfo {year} {2017})}\BibitemShut {NoStop}%
\bibitem [{\citenamefont {Lan}\ \emph {et~al.}(2017)\citenamefont {Lan},
  \citenamefont {van Horssen}, \citenamefont {Powell},\ and\ \citenamefont
  {Garrahan}}]{Lan2017_2}%
  \BibitemOpen
  \bibfield  {author} {\bibinfo {author} {\bibfnamefont {Zhihao}\ \bibnamefont
  {Lan}}, \bibinfo {author} {\bibfnamefont {Merlijn}\ \bibnamefont {van
  Horssen}}, \bibinfo {author} {\bibfnamefont {Stephen}\ \bibnamefont
  {Powell}}, \ and\ \bibinfo {author} {\bibfnamefont {Juan~P}\ \bibnamefont
  {Garrahan}},\ }\bibfield  {title} {\enquote {\bibinfo {title} {Quantum slow
  relaxation and metastability due to dynamical constraints},}\ }\href@noop {}
  {\bibfield  {journal} {\bibinfo  {journal} {arXiv preprint arXiv:1706.02603}\
  } (\bibinfo {year} {2017})}\BibitemShut {NoStop}%
\bibitem [{\citenamefont {Schau{\ss}}\ \emph {et~al.}(2012)\citenamefont
  {Schau{\ss}}, \citenamefont {Cheneau}, \citenamefont {Endres}, \citenamefont
  {Fukuhara}, \citenamefont {Hild}, \citenamefont {Omran}, \citenamefont
  {Pohl}, \citenamefont {Gross}, \citenamefont {Kuhr},\ and\ \citenamefont
  {Bloch}}]{Schauss2012}%
  \BibitemOpen
  \bibfield  {author} {\bibinfo {author} {\bibfnamefont {Peter}\ \bibnamefont
  {Schau{\ss}}}, \bibinfo {author} {\bibfnamefont {Marc}\ \bibnamefont
  {Cheneau}}, \bibinfo {author} {\bibfnamefont {Manuel}\ \bibnamefont
  {Endres}}, \bibinfo {author} {\bibfnamefont {Takeshi}\ \bibnamefont
  {Fukuhara}}, \bibinfo {author} {\bibfnamefont {Sebastian}\ \bibnamefont
  {Hild}}, \bibinfo {author} {\bibfnamefont {Ahmed}\ \bibnamefont {Omran}},
  \bibinfo {author} {\bibfnamefont {Thomas}\ \bibnamefont {Pohl}}, \bibinfo
  {author} {\bibfnamefont {Christian}\ \bibnamefont {Gross}}, \bibinfo {author}
  {\bibfnamefont {Stefan}\ \bibnamefont {Kuhr}}, \ and\ \bibinfo {author}
  {\bibfnamefont {Immanuel}\ \bibnamefont {Bloch}},\ }\bibfield  {title}
  {\enquote {\bibinfo {title} {Observation of spatially ordered structures in a
  two-dimensional rydberg gas},}\ }\href
  {http://dx.doi.org/10.1038/nature11596} {\bibfield  {journal} {\bibinfo
  {journal} {Nature}\ }\textbf {\bibinfo {volume} {491}},\ \bibinfo {pages} {87
  EP --} (\bibinfo {year} {2012})}\BibitemShut {NoStop}%
\bibitem [{\citenamefont {Labuhn}\ \emph {et~al.}(2016)\citenamefont {Labuhn},
  \citenamefont {Barredo}, \citenamefont {Ravets}, \citenamefont
  {de~L{\'e}s{\'e}leuc}, \citenamefont {Macr{\`i}}, \citenamefont {Lahaye},\
  and\ \citenamefont {Browaeys}}]{Labuhn2016}%
  \BibitemOpen
  \bibfield  {author} {\bibinfo {author} {\bibfnamefont {Henning}\ \bibnamefont
  {Labuhn}}, \bibinfo {author} {\bibfnamefont {Daniel}\ \bibnamefont
  {Barredo}}, \bibinfo {author} {\bibfnamefont {Sylvain}\ \bibnamefont
  {Ravets}}, \bibinfo {author} {\bibfnamefont {Sylvain}\ \bibnamefont
  {de~L{\'e}s{\'e}leuc}}, \bibinfo {author} {\bibfnamefont {Tommaso}\
  \bibnamefont {Macr{\`i}}}, \bibinfo {author} {\bibfnamefont {Thierry}\
  \bibnamefont {Lahaye}}, \ and\ \bibinfo {author} {\bibfnamefont {Antoine}\
  \bibnamefont {Browaeys}},\ }\bibfield  {title} {\enquote {\bibinfo {title}
  {Tunable two-dimensional arrays of single rydberg atoms for realizing quantum
  ising models},}\ }\href {http://dx.doi.org/10.1038/nature18274} {\bibfield
  {journal} {\bibinfo  {journal} {Nature}\ }\textbf {\bibinfo {volume} {534}},\
  \bibinfo {pages} {667 EP --} (\bibinfo {year} {2016})}\BibitemShut {NoStop}%
\bibitem [{\citenamefont {{Bernien}}\ \emph {et~al.}(2017)\citenamefont
  {{Bernien}}, \citenamefont {{Schwartz}}, \citenamefont {{Keesling}},
  \citenamefont {{Levine}}, \citenamefont {{Omran}}, \citenamefont {{Pichler}},
  \citenamefont {{Choi}}, \citenamefont {{Zibrov}}, \citenamefont {{Endres}},
  \citenamefont {{Greiner}}, \citenamefont {{Vuleti{\'c}}},\ and\ \citenamefont
  {{Lukin}}}]{Bernien17}%
  \BibitemOpen
  \bibfield  {author} {\bibinfo {author} {\bibfnamefont {H.}~\bibnamefont
  {{Bernien}}}, \bibinfo {author} {\bibfnamefont {S.}~\bibnamefont
  {{Schwartz}}}, \bibinfo {author} {\bibfnamefont {A.}~\bibnamefont
  {{Keesling}}}, \bibinfo {author} {\bibfnamefont {H.}~\bibnamefont
  {{Levine}}}, \bibinfo {author} {\bibfnamefont {A.}~\bibnamefont {{Omran}}},
  \bibinfo {author} {\bibfnamefont {H.}~\bibnamefont {{Pichler}}}, \bibinfo
  {author} {\bibfnamefont {S.}~\bibnamefont {{Choi}}}, \bibinfo {author}
  {\bibfnamefont {A.~S.}\ \bibnamefont {{Zibrov}}}, \bibinfo {author}
  {\bibfnamefont {M.}~\bibnamefont {{Endres}}}, \bibinfo {author}
  {\bibfnamefont {M.}~\bibnamefont {{Greiner}}}, \bibinfo {author}
  {\bibfnamefont {V.}~\bibnamefont {{Vuleti{\'c}}}}, \ and\ \bibinfo {author}
  {\bibfnamefont {M.~D.}\ \bibnamefont {{Lukin}}},\ }\bibfield  {title}
  {\enquote {\bibinfo {title} {{Probing many-body dynamics on a 51-atom quantum
  simulator}},}\ }\href@noop {} {\bibfield  {journal} {\bibinfo  {journal}
  {ArXiv e-prints}\ } (\bibinfo {year} {2017})},\ \Eprint
  {http://arxiv.org/abs/1707.04344} {arXiv:1707.04344 [quant-ph]} \BibitemShut
  {NoStop}%
\bibitem [{\citenamefont {Read}\ and\ \citenamefont
  {Rezayi}(1999)}]{ReadRezayi}%
  \BibitemOpen
  \bibfield  {author} {\bibinfo {author} {\bibfnamefont {N.}~\bibnamefont
  {Read}}\ and\ \bibinfo {author} {\bibfnamefont {E.}~\bibnamefont {Rezayi}},\
  }\bibfield  {title} {\enquote {\bibinfo {title} {Beyond paired quantum hall
  states: Parafermions and incompressible states in the first excited landau
  level},}\ }\href {\doibase 10.1103/PhysRevB.59.8084} {\bibfield  {journal}
  {\bibinfo  {journal} {Phys. Rev. B}\ }\textbf {\bibinfo {volume} {59}},\
  \bibinfo {pages} {8084--8092} (\bibinfo {year} {1999})}\BibitemShut {NoStop}%
\bibitem [{\citenamefont {Nayak}\ \emph {et~al.}(2008)\citenamefont {Nayak},
  \citenamefont {Simon}, \citenamefont {Stern}, \citenamefont {Freedman},\ and\
  \citenamefont {Das~Sarma}}]{NayakRMP}%
  \BibitemOpen
  \bibfield  {author} {\bibinfo {author} {\bibfnamefont {Chetan}\ \bibnamefont
  {Nayak}}, \bibinfo {author} {\bibfnamefont {Steven~H.}\ \bibnamefont
  {Simon}}, \bibinfo {author} {\bibfnamefont {Ady}\ \bibnamefont {Stern}},
  \bibinfo {author} {\bibfnamefont {Michael}\ \bibnamefont {Freedman}}, \ and\
  \bibinfo {author} {\bibfnamefont {Sankar}\ \bibnamefont {Das~Sarma}},\
  }\bibfield  {title} {\enquote {\bibinfo {title} {Non-abelian anyons and
  topological quantum computation},}\ }\href {\doibase
  10.1103/RevModPhys.80.1083} {\bibfield  {journal} {\bibinfo  {journal} {Rev.
  Mod. Phys.}\ }\textbf {\bibinfo {volume} {80}},\ \bibinfo {pages}
  {1083--1159} (\bibinfo {year} {2008})}\BibitemShut {NoStop}%
\bibitem [{\citenamefont {Moudgalya}\ \emph {et~al.}(2017)\citenamefont
  {Moudgalya}, \citenamefont {Rachel}, \citenamefont {Bernevig},\ and\
  \citenamefont {Regnault}}]{Bernevig2017}%
  \BibitemOpen
  \bibfield  {author} {\bibinfo {author} {\bibfnamefont {Sanjay}\ \bibnamefont
  {Moudgalya}}, \bibinfo {author} {\bibfnamefont {Stephan}\ \bibnamefont
  {Rachel}}, \bibinfo {author} {\bibfnamefont {Bogdan~A}\ \bibnamefont
  {Bernevig}}, \ and\ \bibinfo {author} {\bibfnamefont {Nicolas}\ \bibnamefont
  {Regnault}},\ }\bibfield  {title} {\enquote {\bibinfo {title} {Exact excited
  states of non-integrable models},}\ }\href@noop {} {\bibfield  {journal}
  {\bibinfo  {journal} {arXiv preprint arXiv:1708.05021}\ } (\bibinfo {year}
  {2017})}\BibitemShut {NoStop}%
\bibitem [{\citenamefont {Heller}(1984)}]{Heller84}%
  \BibitemOpen
  \bibfield  {author} {\bibinfo {author} {\bibfnamefont {Eric~J.}\ \bibnamefont
  {Heller}},\ }\bibfield  {title} {\enquote {\bibinfo {title} {Bound-state
  eigenfunctions of classically chaotic hamiltonian systems: Scars of periodic
  orbits},}\ }\href {\doibase 10.1103/PhysRevLett.53.1515} {\bibfield
  {journal} {\bibinfo  {journal} {Phys. Rev. Lett.}\ }\textbf {\bibinfo
  {volume} {53}},\ \bibinfo {pages} {1515--1518} (\bibinfo {year}
  {1984})}\BibitemShut {NoStop}%
\bibitem [{\citenamefont {Sridhar}(1991)}]{Sridhar1991}%
  \BibitemOpen
  \bibfield  {author} {\bibinfo {author} {\bibfnamefont {S.}~\bibnamefont
  {Sridhar}},\ }\bibfield  {title} {\enquote {\bibinfo {title} {Experimental
  observation of scarred eigenfunctions of chaotic microwave cavities},}\
  }\href {\doibase 10.1103/PhysRevLett.67.785} {\bibfield  {journal} {\bibinfo
  {journal} {Phys. Rev. Lett.}\ }\textbf {\bibinfo {volume} {67}},\ \bibinfo
  {pages} {785--788} (\bibinfo {year} {1991})}\BibitemShut {NoStop}%
\bibitem [{\citenamefont {Marcus}\ \emph {et~al.}(1992)\citenamefont {Marcus},
  \citenamefont {Rimberg}, \citenamefont {Westervelt}, \citenamefont
  {Hopkins},\ and\ \citenamefont {Gossard}}]{Marcus1992}%
  \BibitemOpen
  \bibfield  {author} {\bibinfo {author} {\bibfnamefont {C.~M.}\ \bibnamefont
  {Marcus}}, \bibinfo {author} {\bibfnamefont {A.~J.}\ \bibnamefont {Rimberg}},
  \bibinfo {author} {\bibfnamefont {R.~M.}\ \bibnamefont {Westervelt}},
  \bibinfo {author} {\bibfnamefont {P.~F.}\ \bibnamefont {Hopkins}}, \ and\
  \bibinfo {author} {\bibfnamefont {A.~C.}\ \bibnamefont {Gossard}},\
  }\bibfield  {title} {\enquote {\bibinfo {title} {Conductance fluctuations and
  chaotic scattering in ballistic microstructures},}\ }\href {\doibase
  10.1103/PhysRevLett.69.506} {\bibfield  {journal} {\bibinfo  {journal} {Phys.
  Rev. Lett.}\ }\textbf {\bibinfo {volume} {69}},\ \bibinfo {pages} {506--509}
  (\bibinfo {year} {1992})}\BibitemShut {NoStop}%
\bibitem [{\citenamefont {Sun}\ and\ \citenamefont
  {Robicheaux}(2008)}]{Sun2008}%
  \BibitemOpen
  \bibfield  {author} {\bibinfo {author} {\bibfnamefont {B}~\bibnamefont
  {Sun}}\ and\ \bibinfo {author} {\bibfnamefont {F}~\bibnamefont
  {Robicheaux}},\ }\bibfield  {title} {\enquote {\bibinfo {title} {Numerical
  study of two-body correlation in a 1d lattice with perfect blockade},}\
  }\href {http://stacks.iop.org/1367-2630/10/i=4/a=045032} {\bibfield
  {journal} {\bibinfo  {journal} {New Journal of Physics}\ }\textbf {\bibinfo
  {volume} {10}},\ \bibinfo {pages} {045032} (\bibinfo {year}
  {2008})}\BibitemShut {NoStop}%
\bibitem [{\citenamefont {Olmos}\ \emph {et~al.}(2009)\citenamefont {Olmos},
  \citenamefont {Gonz\'alez-F\'erez},\ and\ \citenamefont
  {Lesanovsky}}]{LesanovskyDynamics}%
  \BibitemOpen
  \bibfield  {author} {\bibinfo {author} {\bibfnamefont {B.}~\bibnamefont
  {Olmos}}, \bibinfo {author} {\bibfnamefont {R.}~\bibnamefont
  {Gonz\'alez-F\'erez}}, \ and\ \bibinfo {author} {\bibfnamefont
  {I.}~\bibnamefont {Lesanovsky}},\ }\bibfield  {title} {\enquote {\bibinfo
  {title} {Collective rydberg excitations of an atomic gas confined in a ring
  lattice},}\ }\href {\doibase 10.1103/PhysRevA.79.043419} {\bibfield
  {journal} {\bibinfo  {journal} {Phys. Rev. A}\ }\textbf {\bibinfo {volume}
  {79}},\ \bibinfo {pages} {043419} (\bibinfo {year} {2009})}\BibitemShut
  {NoStop}%
\bibitem [{\citenamefont {Vidal}(2007)}]{Vidal07}%
  \BibitemOpen
  \bibfield  {author} {\bibinfo {author} {\bibfnamefont {G.}~\bibnamefont
  {Vidal}},\ }\bibfield  {title} {\enquote {\bibinfo {title} {Classical
  simulation of infinite-size quantum lattice systems in one spatial
  dimension},}\ }\href {\doibase 10.1103/PhysRevLett.98.070201} {\bibfield
  {journal} {\bibinfo  {journal} {Phys. Rev. Lett.}\ }\textbf {\bibinfo
  {volume} {98}},\ \bibinfo {pages} {070201} (\bibinfo {year}
  {2007})}\BibitemShut {NoStop}%
\bibitem [{\citenamefont {Lesanovsky}(2012)}]{LesanovskyMPS}%
  \BibitemOpen
  \bibfield  {author} {\bibinfo {author} {\bibfnamefont {Igor}\ \bibnamefont
  {Lesanovsky}},\ }\bibfield  {title} {\enquote {\bibinfo {title} {Liquid
  ground state, gap, and excited states of a strongly correlated spin chain},}\
  }\href {\doibase 10.1103/PhysRevLett.108.105301} {\bibfield  {journal}
  {\bibinfo  {journal} {Phys. Rev. Lett.}\ }\textbf {\bibinfo {volume} {108}},\
  \bibinfo {pages} {105301} (\bibinfo {year} {2012})}\BibitemShut {NoStop}%
\bibitem [{\citenamefont {Lanczos}(1950)}]{Lanczos}%
  \BibitemOpen
  \bibfield  {author} {\bibinfo {author} {\bibfnamefont {Cornelius}\
  \bibnamefont {Lanczos}},\ }\href@noop {} {\emph {\bibinfo {title} {An
  iteration method for the solution of the eigenvalue problem of linear
  differential and integral operators}}}\ (\bibinfo  {publisher} {United States
  Governm. Press Office Los Angeles, CA},\ \bibinfo {year} {1950})\BibitemShut
  {NoStop}%
\bibitem [{Note1()}]{Note1}%
  \BibitemOpen
  \bibinfo {note} {A. Michailidis \protect \emph {et al.}, in
  preparation.}\BibitemShut {Stop}%
\bibitem [{\citenamefont {Fendley}\ \emph {et~al.}(2004)\citenamefont
  {Fendley}, \citenamefont {Sengupta},\ and\ \citenamefont
  {Sachdev}}]{FendleySachdev}%
  \BibitemOpen
  \bibfield  {author} {\bibinfo {author} {\bibfnamefont {Paul}\ \bibnamefont
  {Fendley}}, \bibinfo {author} {\bibfnamefont {K.}~\bibnamefont {Sengupta}}, \
  and\ \bibinfo {author} {\bibfnamefont {Subir}\ \bibnamefont {Sachdev}},\
  }\bibfield  {title} {\enquote {\bibinfo {title} {Competing density-wave
  orders in a one-dimensional hard-boson model},}\ }\href {\doibase
  10.1103/PhysRevB.69.075106} {\bibfield  {journal} {\bibinfo  {journal} {Phys.
  Rev. B}\ }\textbf {\bibinfo {volume} {69}},\ \bibinfo {pages} {075106}
  (\bibinfo {year} {2004})}\BibitemShut {NoStop}%
\bibitem [{\citenamefont {Fendley}(2016)}]{FendleyXYZ}%
  \BibitemOpen
  \bibfield  {author} {\bibinfo {author} {\bibfnamefont {Paul}\ \bibnamefont
  {Fendley}},\ }\bibfield  {title} {\enquote {\bibinfo {title} {Strong zero
  modes and eigenstate phase transitions in the xyz/interacting majorana
  chain},}\ }\href {http://stacks.iop.org/1751-8121/49/i=30/a=30LT01}
  {\bibfield  {journal} {\bibinfo  {journal} {Journal of Physics A:
  Mathematical and Theoretical}\ }\textbf {\bibinfo {volume} {49}},\ \bibinfo
  {pages} {30LT01} (\bibinfo {year} {2016})}\BibitemShut {NoStop}%
\bibitem [{\citenamefont {Bogomolny}\ \emph {et~al.}(1999)\citenamefont
  {Bogomolny}, \citenamefont {Gerland},\ and\ \citenamefont
  {Schmit}}]{Bogomolny99}%
  \BibitemOpen
  \bibfield  {author} {\bibinfo {author} {\bibfnamefont {E.~B.}\ \bibnamefont
  {Bogomolny}}, \bibinfo {author} {\bibfnamefont {U.}~\bibnamefont {Gerland}},
  \ and\ \bibinfo {author} {\bibfnamefont {C.}~\bibnamefont {Schmit}},\
  }\bibfield  {title} {\enquote {\bibinfo {title} {Models of intermediate
  spectral statistics},}\ }\href {\doibase 10.1103/PhysRevE.59.R1315}
  {\bibfield  {journal} {\bibinfo  {journal} {Phys. Rev. E}\ }\textbf {\bibinfo
  {volume} {59}},\ \bibinfo {pages} {R1315--R1318} (\bibinfo {year}
  {1999})}\BibitemShut {NoStop}%
\bibitem [{\citenamefont {Sutherland}(1986)}]{Sutherland86}%
  \BibitemOpen
  \bibfield  {author} {\bibinfo {author} {\bibfnamefont {Bill}\ \bibnamefont
  {Sutherland}},\ }\bibfield  {title} {\enquote {\bibinfo {title} {Localization
  of electronic wave functions due to local topology},}\ }\href {\doibase
  10.1103/PhysRevB.34.5208} {\bibfield  {journal} {\bibinfo  {journal} {Phys.
  Rev. B}\ }\textbf {\bibinfo {volume} {34}},\ \bibinfo {pages} {5208--5211}
  (\bibinfo {year} {1986})}\BibitemShut {NoStop}%
\bibitem [{\citenamefont {Inui}\ \emph {et~al.}(1994)\citenamefont {Inui},
  \citenamefont {Trugman},\ and\ \citenamefont {Abrahams}}]{Inui}%
  \BibitemOpen
  \bibfield  {author} {\bibinfo {author} {\bibfnamefont {M.}~\bibnamefont
  {Inui}}, \bibinfo {author} {\bibfnamefont {S.~A.}\ \bibnamefont {Trugman}}, \
  and\ \bibinfo {author} {\bibfnamefont {Elihu}\ \bibnamefont {Abrahams}},\
  }\bibfield  {title} {\enquote {\bibinfo {title} {Unusual properties of
  midband states in systems with off-diagonal disorder},}\ }\href {\doibase
  10.1103/PhysRevB.49.3190} {\bibfield  {journal} {\bibinfo  {journal} {Phys.
  Rev. B}\ }\textbf {\bibinfo {volume} {49}},\ \bibinfo {pages} {3190--3196}
  (\bibinfo {year} {1994})}\BibitemShut {NoStop}%
\bibitem [{\citenamefont {Sutherland}(1971)}]{CalogeroSutherland}%
  \BibitemOpen
  \bibfield  {author} {\bibinfo {author} {\bibfnamefont {Bill}\ \bibnamefont
  {Sutherland}},\ }\bibfield  {title} {\enquote {\bibinfo {title} {Exact
  results for a quantum many-body problem in one dimension},}\ }\href {\doibase
  10.1103/PhysRevA.4.2019} {\bibfield  {journal} {\bibinfo  {journal} {Phys.
  Rev. A}\ }\textbf {\bibinfo {volume} {4}},\ \bibinfo {pages} {2019--2021}
  (\bibinfo {year} {1971})}\BibitemShut {NoStop}%
\bibitem [{\citenamefont {Bernevig}\ and\ \citenamefont
  {Haldane}(2008)}]{BernevigHaldane}%
  \BibitemOpen
  \bibfield  {author} {\bibinfo {author} {\bibfnamefont {B.~Andrei}\
  \bibnamefont {Bernevig}}\ and\ \bibinfo {author} {\bibfnamefont {F.~D.~M.}\
  \bibnamefont {Haldane}},\ }\bibfield  {title} {\enquote {\bibinfo {title}
  {Model fractional quantum hall states and jack polynomials},}\ }\href
  {\doibase 10.1103/PhysRevLett.100.246802} {\bibfield  {journal} {\bibinfo
  {journal} {Phys. Rev. Lett.}\ }\textbf {\bibinfo {volume} {100}},\ \bibinfo
  {pages} {246802} (\bibinfo {year} {2008})}\BibitemShut {NoStop}%
\bibitem [{\citenamefont {Carleo}\ \emph {et~al.}(2012)\citenamefont {Carleo},
  \citenamefont {Becca}, \citenamefont {Schir{\'o}},\ and\ \citenamefont
  {Fabrizio}}]{carleo2012localization}%
  \BibitemOpen
  \bibfield  {author} {\bibinfo {author} {\bibfnamefont {Giuseppe}\
  \bibnamefont {Carleo}}, \bibinfo {author} {\bibfnamefont {Federico}\
  \bibnamefont {Becca}}, \bibinfo {author} {\bibfnamefont {Marco}\ \bibnamefont
  {Schir{\'o}}}, \ and\ \bibinfo {author} {\bibfnamefont {Michele}\
  \bibnamefont {Fabrizio}},\ }\bibfield  {title} {\enquote {\bibinfo {title}
  {Localization and glassy dynamics of many-body quantum systems},}\ }\href
  {http://dx.doi.org/10.1038/srep00243} {\bibfield  {journal} {\bibinfo
  {journal} {Scientific Reports}\ }\textbf {\bibinfo {volume} {2}},\ \bibinfo
  {pages} {243} (\bibinfo {year} {2012})}\BibitemShut {NoStop}%
\bibitem [{\citenamefont {De~Roeck}\ and\ \citenamefont
  {Huveneers}(2014{\natexlab{a}})}]{Huveneers13}%
  \BibitemOpen
  \bibfield  {author} {\bibinfo {author} {\bibfnamefont {Wojciech}\
  \bibnamefont {De~Roeck}}\ and\ \bibinfo {author} {\bibfnamefont {Fran{\c
  c}ois}\ \bibnamefont {Huveneers}},\ }\bibfield  {title} {{\selectlanguage
  {English}\enquote {\bibinfo {title} {Asymptotic quantum many-body
  localization from thermal disorder},}\ }}\href {\doibase
  10.1007/s00220-014-2116-8} {\bibfield  {journal} {\bibinfo  {journal}
  {Communications in Mathematical Physics}\ }\textbf {\bibinfo {volume}
  {332}},\ \bibinfo {pages} {1017--1082} (\bibinfo {year}
  {2014}{\natexlab{a}})}\BibitemShut {NoStop}%
\bibitem [{\citenamefont {{Schiulaz}}\ and\ \citenamefont
  {{M{\"u}ller}}(2014)}]{Muller}%
  \BibitemOpen
  \bibfield  {author} {\bibinfo {author} {\bibfnamefont {M.}~\bibnamefont
  {{Schiulaz}}}\ and\ \bibinfo {author} {\bibfnamefont {M.}~\bibnamefont
  {{M{\"u}ller}}},\ }\bibfield  {title} {\enquote {\bibinfo {title} {{Ideal
  quantum glass transitions: Many-body localization without quenched
  disorder}},}\ }in\ \href {\doibase 10.1063/1.4893505} {\emph {\bibinfo
  {booktitle} {American Institute of Physics Conference Series}}},\ \bibinfo
  {series} {American Institute of Physics Conference Series}, Vol.\ \bibinfo
  {volume} {1610}\ (\bibinfo {year} {2014})\ pp.\ \bibinfo {pages} {11--23},\
  \Eprint {http://arxiv.org/abs/1309.1082} {arXiv:1309.1082 [cond-mat.dis-nn]}
  \BibitemShut {NoStop}%
\bibitem [{\citenamefont {Grover}\ and\ \citenamefont {Fisher}(2014)}]{QDL}%
  \BibitemOpen
  \bibfield  {author} {\bibinfo {author} {\bibfnamefont {Tarun}\ \bibnamefont
  {Grover}}\ and\ \bibinfo {author} {\bibfnamefont {Matthew P~A}\ \bibnamefont
  {Fisher}},\ }\bibfield  {title} {\enquote {\bibinfo {title} {Quantum
  disentangled liquids},}\ }\href
  {http://stacks.iop.org/1742-5468/2014/i=10/a=P10010} {\bibfield  {journal}
  {\bibinfo  {journal} {Journal of Statistical Mechanics: Theory and
  Experiment}\ }\textbf {\bibinfo {volume} {2014}},\ \bibinfo {pages} {P10010}
  (\bibinfo {year} {2014})}\BibitemShut {NoStop}%
\bibitem [{\citenamefont {Yao}\ \emph {et~al.}(2016)\citenamefont {Yao},
  \citenamefont {Laumann}, \citenamefont {Cirac}, \citenamefont {Lukin},\ and\
  \citenamefont {Moore}}]{Yao14}%
  \BibitemOpen
  \bibfield  {author} {\bibinfo {author} {\bibfnamefont {N.~Y.}\ \bibnamefont
  {Yao}}, \bibinfo {author} {\bibfnamefont {C.~R.}\ \bibnamefont {Laumann}},
  \bibinfo {author} {\bibfnamefont {J.~I.}\ \bibnamefont {Cirac}}, \bibinfo
  {author} {\bibfnamefont {M.~D.}\ \bibnamefont {Lukin}}, \ and\ \bibinfo
  {author} {\bibfnamefont {J.~E.}\ \bibnamefont {Moore}},\ }\bibfield  {title}
  {\enquote {\bibinfo {title} {Quasi-many-body localization in
  translation-invariant systems},}\ }\href {\doibase
  10.1103/PhysRevLett.117.240601} {\bibfield  {journal} {\bibinfo  {journal}
  {Phys. Rev. Lett.}\ }\textbf {\bibinfo {volume} {117}},\ \bibinfo {pages}
  {240601} (\bibinfo {year} {2016})}\BibitemShut {NoStop}%
\bibitem [{\citenamefont {Garrison}\ \emph {et~al.}(2017)\citenamefont
  {Garrison}, \citenamefont {Mishmash},\ and\ \citenamefont {Fisher}}]{QDL2}%
  \BibitemOpen
  \bibfield  {author} {\bibinfo {author} {\bibfnamefont {James~R.}\
  \bibnamefont {Garrison}}, \bibinfo {author} {\bibfnamefont {Ryan~V.}\
  \bibnamefont {Mishmash}}, \ and\ \bibinfo {author} {\bibfnamefont {Matthew
  P.~A.}\ \bibnamefont {Fisher}},\ }\bibfield  {title} {\enquote {\bibinfo
  {title} {Partial breakdown of quantum thermalization in a hubbard-like
  model},}\ }\href {\doibase 10.1103/PhysRevB.95.054204} {\bibfield  {journal}
  {\bibinfo  {journal} {Phys. Rev. B}\ }\textbf {\bibinfo {volume} {95}},\
  \bibinfo {pages} {054204} (\bibinfo {year} {2017})}\BibitemShut {NoStop}%
\bibitem [{\citenamefont {Veness}\ \emph {et~al.}(2016)\citenamefont {Veness},
  \citenamefont {Essler},\ and\ \citenamefont {Fisher}}]{QDLEssler}%
  \BibitemOpen
  \bibfield  {author} {\bibinfo {author} {\bibfnamefont {Thomas}\ \bibnamefont
  {Veness}}, \bibinfo {author} {\bibfnamefont {Fabian H.~L.}\ \bibnamefont
  {Essler}}, \ and\ \bibinfo {author} {\bibfnamefont {M.~P.~A.}\ \bibnamefont
  {Fisher}},\ }\bibfield  {title} {\enquote {\bibinfo {title} {{Disorder-free
  localization}},}\ }\href@noop {} {\bibfield  {journal} {\bibinfo  {journal}
  {ArXiv e-prints}\ } (\bibinfo {year} {2016})},\ \Eprint
  {http://arxiv.org/abs/1611.02075} {arXiv:1611.02075 [cond-mat.dis-nn]}
  \BibitemShut {NoStop}%
\bibitem [{\citenamefont {De~Roeck}\ and\ \citenamefont
  {Huveneers}(2014{\natexlab{b}})}]{PhysRevB.90.165137}%
  \BibitemOpen
  \bibfield  {author} {\bibinfo {author} {\bibfnamefont {Wojciech}\
  \bibnamefont {De~Roeck}}\ and\ \bibinfo {author} {\bibfnamefont
  {Fran\c{c}ois}\ \bibnamefont {Huveneers}},\ }\bibfield  {title} {\enquote
  {\bibinfo {title} {Scenario for delocalization in translation-invariant
  systems},}\ }\href {\doibase 10.1103/PhysRevB.90.165137} {\bibfield
  {journal} {\bibinfo  {journal} {Phys. Rev. B}\ }\textbf {\bibinfo {volume}
  {90}},\ \bibinfo {pages} {165137} (\bibinfo {year}
  {2014}{\natexlab{b}})}\BibitemShut {NoStop}%
\bibitem [{\citenamefont {Kim}\ and\ \citenamefont {Haah}(2016)}]{Kim2016}%
  \BibitemOpen
  \bibfield  {author} {\bibinfo {author} {\bibfnamefont {Isaac~H.}\
  \bibnamefont {Kim}}\ and\ \bibinfo {author} {\bibfnamefont {Jeongwan}\
  \bibnamefont {Haah}},\ }\bibfield  {title} {\enquote {\bibinfo {title}
  {Localization from superselection rules in translationally invariant
  systems},}\ }\href {\doibase 10.1103/PhysRevLett.116.027202} {\bibfield
  {journal} {\bibinfo  {journal} {Phys. Rev. Lett.}\ }\textbf {\bibinfo
  {volume} {116}},\ \bibinfo {pages} {027202} (\bibinfo {year}
  {2016})}\BibitemShut {NoStop}%
\bibitem [{\citenamefont {Yarloo}\ \emph {et~al.}(2017)\citenamefont {Yarloo},
  \citenamefont {Langari},\ and\ \citenamefont {Vaezi}}]{Yarloo2017}%
  \BibitemOpen
  \bibfield  {author} {\bibinfo {author} {\bibfnamefont {H.}~\bibnamefont
  {Yarloo}}, \bibinfo {author} {\bibfnamefont {A.}~\bibnamefont {Langari}}, \
  and\ \bibinfo {author} {\bibfnamefont {A.}~\bibnamefont {Vaezi}},\ }\bibfield
   {title} {\enquote {\bibinfo {title} {Anyonic self-induced disorder in a
  stabilizer code: many body localization in a translational invariant
  model},}\ }\href@noop {} {\bibfield  {journal} {\bibinfo  {journal} {arXiv
  preprint arXiv:1703.06621}\ } (\bibinfo {year} {2017})}\BibitemShut {NoStop}%
\bibitem [{\citenamefont {Michailidis}\ \emph {et~al.}(2017)\citenamefont
  {Michailidis}, \citenamefont {{\v{Z}}nidari{\v{c}}}, \citenamefont
  {Medvedyeva}, \citenamefont {Abanin}, \citenamefont {Prosen},\ and\
  \citenamefont {Papi{\'c}}}]{michailidis2017slow}%
  \BibitemOpen
  \bibfield  {author} {\bibinfo {author} {\bibfnamefont {Alexios~A}\
  \bibnamefont {Michailidis}}, \bibinfo {author} {\bibfnamefont {Marko}\
  \bibnamefont {{\v{Z}}nidari{\v{c}}}}, \bibinfo {author} {\bibfnamefont
  {Mariya}\ \bibnamefont {Medvedyeva}}, \bibinfo {author} {\bibfnamefont
  {Dmitry~A}\ \bibnamefont {Abanin}}, \bibinfo {author} {\bibfnamefont
  {Toma{\v{z}}}\ \bibnamefont {Prosen}}, \ and\ \bibinfo {author}
  {\bibfnamefont {Z}~\bibnamefont {Papi{\'c}}},\ }\bibfield  {title} {\enquote
  {\bibinfo {title} {Slow dynamics in translation-invariant quantum lattice
  models},}\ }\href@noop {} {\bibfield  {journal} {\bibinfo  {journal} {arXiv
  preprint arXiv:1706.05026}\ } (\bibinfo {year} {2017})}\BibitemShut {NoStop}%
\bibitem [{\citenamefont {Smith}\ \emph {et~al.}(2017)\citenamefont {Smith},
  \citenamefont {Knolle}, \citenamefont {Moessner},\ and\ \citenamefont
  {Kovrizhin}}]{Smith2017}%
  \BibitemOpen
  \bibfield  {author} {\bibinfo {author} {\bibfnamefont {A.}~\bibnamefont
  {Smith}}, \bibinfo {author} {\bibfnamefont {J.}~\bibnamefont {Knolle}},
  \bibinfo {author} {\bibfnamefont {R.}~\bibnamefont {Moessner}}, \ and\
  \bibinfo {author} {\bibfnamefont {D.~L.}\ \bibnamefont {Kovrizhin}},\
  }\bibfield  {title} {\enquote {\bibinfo {title} {Absence of ergodicity
  without quenched disorder: From quantum disentangled liquids to many-body
  localization},}\ }\href {\doibase 10.1103/PhysRevLett.119.176601} {\bibfield
  {journal} {\bibinfo  {journal} {Phys. Rev. Lett.}\ }\textbf {\bibinfo
  {volume} {119}},\ \bibinfo {pages} {176601} (\bibinfo {year}
  {2017})}\BibitemShut {NoStop}%
\bibitem [{\citenamefont {Brenes}\ \emph {et~al.}(2017)\citenamefont {Brenes},
  \citenamefont {Dalmonte}, \citenamefont {Heyl},\ and\ \citenamefont
  {Scardicchio}}]{brenes2017many}%
  \BibitemOpen
  \bibfield  {author} {\bibinfo {author} {\bibfnamefont {Marlon}\ \bibnamefont
  {Brenes}}, \bibinfo {author} {\bibfnamefont {Marcello}\ \bibnamefont
  {Dalmonte}}, \bibinfo {author} {\bibfnamefont {Markus}\ \bibnamefont {Heyl}},
  \ and\ \bibinfo {author} {\bibfnamefont {Antonello}\ \bibnamefont
  {Scardicchio}},\ }\bibfield  {title} {\enquote {\bibinfo {title} {Many-body
  localization dynamics from gauge invariance},}\ }\href@noop {} {\bibfield
  {journal} {\bibinfo  {journal} {arXiv preprint arXiv:1706.05878}\ } (\bibinfo
  {year} {2017})}\BibitemShut {NoStop}%
\bibitem [{\citenamefont {{Garrahan}}\ \emph {et~al.}(2010)\citenamefont
  {{Garrahan}}, \citenamefont {{Sollich}},\ and\ \citenamefont
  {{Toninelli}}}]{KCM-rev}%
  \BibitemOpen
  \bibfield  {author} {\bibinfo {author} {\bibfnamefont {J.~P.}\ \bibnamefont
  {{Garrahan}}}, \bibinfo {author} {\bibfnamefont {P.}~\bibnamefont
  {{Sollich}}}, \ and\ \bibinfo {author} {\bibfnamefont {C.}~\bibnamefont
  {{Toninelli}}},\ }\bibfield  {title} {\enquote {\bibinfo {title}
  {{Kinetically Constrained Models}},}\ }\href@noop {} {\bibfield  {journal}
  {\bibinfo  {journal} {ArXiv e-prints}\ } (\bibinfo {year} {2010})},\ \Eprint
  {http://arxiv.org/abs/1009.6113} {arXiv:1009.6113 [cond-mat.stat-mech]}
  \BibitemShut {NoStop}%
\bibitem [{\citenamefont {van Horssen}\ \emph {et~al.}(2015)\citenamefont {van
  Horssen}, \citenamefont {Levi},\ and\ \citenamefont {Garrahan}}]{Juan15}%
  \BibitemOpen
  \bibfield  {author} {\bibinfo {author} {\bibfnamefont {Merlijn}\ \bibnamefont
  {van Horssen}}, \bibinfo {author} {\bibfnamefont {Emanuele}\ \bibnamefont
  {Levi}}, \ and\ \bibinfo {author} {\bibfnamefont {Juan~P.}\ \bibnamefont
  {Garrahan}},\ }\bibfield  {title} {\enquote {\bibinfo {title} {Dynamics of
  many-body localization in a translation-invariant quantum glass model},}\
  }\href {\doibase 10.1103/PhysRevB.92.100305} {\bibfield  {journal} {\bibinfo
  {journal} {Phys. Rev. B}\ }\textbf {\bibinfo {volume} {92}},\ \bibinfo
  {pages} {100305} (\bibinfo {year} {2015})}\BibitemShut {NoStop}%
\bibitem [{\citenamefont {Hickey}\ \emph {et~al.}(2016)\citenamefont {Hickey},
  \citenamefont {Genway},\ and\ \citenamefont {Garrahan}}]{Juan16}%
  \BibitemOpen
  \bibfield  {author} {\bibinfo {author} {\bibfnamefont {James~M}\ \bibnamefont
  {Hickey}}, \bibinfo {author} {\bibfnamefont {Sam}\ \bibnamefont {Genway}}, \
  and\ \bibinfo {author} {\bibfnamefont {Juan~P}\ \bibnamefont {Garrahan}},\
  }\bibfield  {title} {\enquote {\bibinfo {title} {Signatures of many-body
  localisation in a system without disorder and the relation to a glass
  transition},}\ }\href {http://stacks.iop.org/1742-5468/2016/i=5/a=054047}
  {\bibfield  {journal} {\bibinfo  {journal} {Journal of Statistical Mechanics:
  Theory and Experiment}\ }\textbf {\bibinfo {volume} {2016}},\ \bibinfo
  {pages} {054047} (\bibinfo {year} {2016})}\BibitemShut {NoStop}%
\bibitem [{\citenamefont {Bloch}\ \emph {et~al.}(2012)\citenamefont {Bloch},
  \citenamefont {Dalibard},\ and\ \citenamefont {Nascimb{\`e}ne}}]{Bloch2012}%
  \BibitemOpen
  \bibfield  {author} {\bibinfo {author} {\bibfnamefont {Immanuel}\
  \bibnamefont {Bloch}}, \bibinfo {author} {\bibfnamefont {Jean}\ \bibnamefont
  {Dalibard}}, \ and\ \bibinfo {author} {\bibfnamefont {Sylvain}\ \bibnamefont
  {Nascimb{\`e}ne}},\ }\bibfield  {title} {\enquote {\bibinfo {title} {Quantum
  simulations with ultracold quantum gases},}\ }\href
  {http://dx.doi.org/10.1038/nphys2259} {\bibfield  {journal} {\bibinfo
  {journal} {Nature Physics}\ }\textbf {\bibinfo {volume} {8}},\ \bibinfo
  {pages} {267 EP --} (\bibinfo {year} {2012})},\ \bibinfo {note} {review
  Article}\BibitemShut {NoStop}%
\bibitem [{\citenamefont {Blatt}\ and\ \citenamefont {Roos}(2012)}]{Blatt2012}%
  \BibitemOpen
  \bibfield  {author} {\bibinfo {author} {\bibfnamefont {R.}~\bibnamefont
  {Blatt}}\ and\ \bibinfo {author} {\bibfnamefont {C.~F.}\ \bibnamefont
  {Roos}},\ }\bibfield  {title} {\enquote {\bibinfo {title} {Quantum
  simulations with trapped ions},}\ }\href
  {http://dx.doi.org/10.1038/nphys2252} {\bibfield  {journal} {\bibinfo
  {journal} {Nature Physics}\ }\textbf {\bibinfo {volume} {8}},\ \bibinfo
  {pages} {277 EP --} (\bibinfo {year} {2012})},\ \bibinfo {note} {review
  Article}\BibitemShut {NoStop}%
\end{thebibliography}%

\end{document}